\documentclass{article}
 \usepackage{mathtools}
 \usepackage{ mathrsfs }

\usepackage{amssymb}
\usepackage{amscd}
\usepackage{bm}
\usepackage{graphics}
\usepackage{amsfonts}

\usepackage{amsthm}
\usepackage{amsmath}
\usepackage{latexsym}
\usepackage[all]{xy}
\usepackage{fancyhdr}
\usepackage{color}
\usepackage{cite}

\usepackage{latexsym,graphics,color}

\def\be{\begin{equation}}
\def\ee{\end{equation}}

\def\D{\mathcal D}

\def\ka{\kappa}

\def\P{\mathcal P}

\def\E{\mathcal E}
\def\S{\mathcal S}
\def\F{\mathcal F}
\def\A{\mathcal A}
\def\bc{\mathbb C}
\def\br{\mathbb R}
 \def\tr{{\rm tr}}

\def\K{\mathcal K}
\def\J{\mathcal J}
\def\L{\mathcal L}

\def\js{\frac{1}{4}}
\def\ad{{\rm ad}}

\def\d{\partial}
\def\jp{\frac{1}{2}}
\def\rjp{\frac{\rho_R}{2}}
\def\ljp{\frac{\rho_L}{2}}
\def\ri{{\mathrm i}}

\def\al{\alpha}

\def\col{\cos{\left(\frac{\rho_L}{2}\right)}}
\def\sir{\sin{\left(\frac{\rho_R}{2}\right)}}
\def\cor{\cos{\left(\frac{\rho_R}{2}\right)}}

\def\cN{{\cal N}}     %
\def\ri{{\mathrm{i}}}                   %
\def\1{{\mbox{\boldmath $1$}}}          %
\def\tr{\mathrm{tr\,}}                  %
\def\e{\epsilon}                        %

\def\vk{\varkappa}%
\def\jp{\frac{1}{2}}                    %
\def\al{\alpha}                         %

\definecolor{spec}{rgb}{0.0, 0.26, 0.15}

\def\Ad{{\rm Ad}}
\def\bm{\begin{pmatrix}}
\def\em{\end{pmatrix}}

  \def\sr{\sin{\rjp}}
    \def\sl{\sin{\ljp}}
      \def\cor{\cos{\rjp}}
        \def\col{\cos{\ljp}}
  \def\tar{\tan{\rjp}}
    \def\tal{\tan{\ljp}}

\begin{document}

\begin{flushright}
{}~
  
\end{flushright}

\vspace{1cm}
\begin{center}
{\large \bf Dressing cosets and multi-parametric integrable deformations}

\vspace{1cm}

{\small
{\bf Ctirad Klim\v{c}\'{\i}k}
\\
Aix Marseille Universit\'e, CNRS, Centrale Marseille\\ I2M, UMR 7373\\ 13453 Marseille, France}
\end{center}

\vspace{0.5 cm}

\centerline{\bf Abstract}
\vspace{0.5 cm}
\noindent   We provide a new construction  of the dressing cosets $\sigma$-models which is based on an isotropic gauging of the $\E$-models. As an application of this new approach, we show that the recently constructed multi-parametric integrable deformations of the principal chiral model are the dressing cosets, they are therefore automatically renormalizable and their dynamics can be completely characterised in terms of  current algebras.

\vspace{2pc}


\vspace{2pc}
 
 \section{Summary of the results}

 \setcounter{equation}{0}
 
 In this paper, we study the  integrable $\sigma$-model which was recently proposed in \cite{DHKM} by Delduc, Hoare, Kameyama and Magro (DHKM)  as the generalisation of the Lukyanov model \cite{L12}. This DHKM model lives
 on the simple compact group target $K$ and its action reads 
 $$S(k)= \kappa I_{\rm WZ}(k)-$$\be  -\jp \int d\tau \oint \tr \biggl[ k^{-1}\partial_+k \biggl(\Bigl(1-\kappa(Q_{L}-Q_R)\Bigr)\Bigl(Q_{L}+Q_R\Bigr)^{-1}+\Big(P_{L}+P_R\Bigr)^{-1}\Bigl(1+\kappa(P_{L}-P_R)\Bigr)\biggr)k^{-1}\partial_-k\biggr].\label{438}\ee
Here $\tau$ and $\sigma$ are, respectively, the worldsheet time and (compact) space variable, $\oint$ stands for the integral over the angle $\sigma$ and the light-cone derivatives are defined as $\partial_\pm=\partial_\tau\pm\partial_\sigma$. Furthermore, $k=k(\tau,\sigma)$  is a $K$-valued  field,  the standard Wess-Zumino term $I_{\rm WZ}$ is defined as 
\be I_{\rm WZ}(k)= -\int d^{-1}\oint \tr\left(dkk^{-1}\wedge[\partial_\sigma kk^{-1}, dkk^{-1}]\right)\label{440}\ee
and $Q_{L,R}$ as well as $P_{L,R}$ are field-dependent $\br$-linear operators on the Lie algebra $\K$ constructed out of the  field-independent $\br$-linear operators  $M_{LL},M_{LR},M_{RL},M_{RR}:\K\to\K$   as follows 
$$ Q_L:=\Ad_{k^{-1}}\Bigl(M_{LL}+\kappa - M_{LR}(M_{RR}-\kappa)^{-1}M_{RL}\Bigr)^{-1}\Bigl(Ad_k+M_{LR}(M_{RR}-\kappa)^{-1} \Bigr) ,$$$$
  Q_R:=\Bigl(M_{RR}-\kappa- M_{RL}(M_{LL}+\kappa)^{-1}M_{LR}\Bigr)^{-1}\Bigl(1+M_{RL}(M_{LL}+\kappa)^{-1}\Ad_k\Bigr),$$ 
$$P_L:=\Bigl(\Ad_{k^{-1}}+(M_{RR}+\kappa)^{-1}M_{RL}\Bigr)\Bigl(M_{LL}-\kappa- M_{LR}(M_{RR}+\kappa)^{-1}M_{RL}\Bigr)^{-1}\Ad_k, $$\be P_R:= \Bigl(1+\Ad_{k^{-1}}(M_{LL}-\kappa)^{-1}M_{LR}\Bigr)\Bigl(M_{RR}+\kappa- M_{RL}(M_{LL}-\kappa)^{-1}M_{LR}\Bigr)^{-1}. \label{446}\ee
The operators  $M_{LL},M_{LR},M_{RL},M_{RR}$ themselves are given by
  $$M_{LL}=\biggl(1+\A_LR+\Bigl(\eta_L^2 -(1+\eta_R^2)(1+\eta_L^2+\kappa)(1+\eta_L^2-\kappa)\omega^t\tilde \cN^{-1}\omega\Bigr)(R^2+1)\biggr),$$
 $$M_{RR}=\biggl(1+\A_RR+\Bigl(\eta_R^2 -(1+\eta_L^2)(1+\eta_R^2+\kappa)(1+\eta_R^2-\kappa)\omega  \cN^{-1}\omega^t\Bigr)(R^2+1)\biggr),$$
 $$ M_{LR}=(1+\eta_L^2+\kappa)(1+\eta_R^2+\kappa)\cN^{-1}\omega^t(R^2+1), $$\be
 M_{RL}=-(1+\eta_L^2-\kappa)(1+\eta_R^2-\kappa)\tilde\cN^{-1}\omega(R^2+1),\label{484}\ee
where
  \be \cN=1+(1+\eta_L^2)(1+\eta_R)^2\omega^t\omega,\quad \tilde\cN=1+(1+\eta_L^2)(1+\eta_R)^2\omega\omega^t,\label{621}\ee
  \be \A_L^2=\eta_L^2\left(1-\frac{\kappa^2}{1+\eta_L^2}\right),\quad
 \A_R^2=\eta_R^2\left(1-\frac{\kappa^2}{1+\eta_R^2}\right).\label{453}\ee
Moreover, $R:\K\to\K$ is the Yang-Baxter $\br$-linear operator which annihilates the Cartan subalgebra
 of $\K$ and it is defined as \be RB^\al=C^\al, \quad RC^\al=-B^\al,\ee
 where $B^\al,C^\al$ are given in terms of the  step generators of $\K^\bc$ as \be B^\al=\frac{\ri}{\sqrt{2}}(E^\al+E^{-\al}), \quad C^\al=\frac{1}{\sqrt{2}}(E^\al-E^{-\al}).\ee 
 Finally, the independent parameters characterizing the DHKM model are: the positive integer   $\kappa$, two real numbers $\eta_L,\eta_R$ such that $\eta_a^2\geq \ka^2-1$, $a=L,R$ and one real $r\times r$ matrix $\omega$ the entries of which are called TsT parameters. Here  $r$ is the dimension of the Cartan torus of the group $K$ and, in the formula above, $\omega:\mathcal{T}\to\mathcal{T}$ is understood as the $\br$-linear operator acting on the Cartan subalgebra $\mathcal{T} \subset\K$.  
 
 \smallskip
 
 Here is the list of the original results obtained in the present article:
 
 \smallskip
 
 \noindent 1) We show that the presence of the TsT parameter matrix $\omega$ in the Lagrangian has no impact neither  on the first order Hamiltonian dynamics of the DHKM model nor on its renormalizability, although it is true at the same time that this presence  does influence the target space geometry. Said in other words, we show that the models with different 
 TsT matrices $\omega_1\neq \omega_2$ are T-dual to each other (the T-duality in question turns out to be the Poisson-Lie T-duality \cite{KS95}) therefore for the understanding of the Hamiltonian dynamics of the DHKM model  and of its renormalizability it is fully sufficient to consider the simplest case $\omega=0$.
 
 \smallskip
 
 \noindent 2) In the case $\omega=0$, we succeed   to rewrite the DHKM action in the following  compact form  
  \be S(k)= \kappa I_{\rm WZ}(k)  + \kappa \int d\tau \oint \  \tr k^{-1}\partial_+k\ \! \frac{\al+ e^{\rho_LR_k}e^{\rho_RR}}{\al- e^{\rho_LR_k}e^{\rho_RR}} \ \! k^{-1}\partial_-k.\label{210c}\ee
Here $R$ is the Yang-Baxter operator, 
 $R_k$ stands for the operator Ad$_{k^{-1}}R$Ad$_k$ and the parameters $\kappa,\al,\rho_L,\rho_R$ are related to the original DHKM parameters $\kappa,\eta_L,\eta_R$
 as follows
 \be \kappa=\kappa, \qquad \alpha=\frac{\eta_L^2+1-\ka}{\eta_L^2+1+\ka}\ \!\frac{\eta_R^2+1-\ka}{\eta_R^2+1+\ka},\qquad  \tan{\frac{\rho_a}{2}}=\frac{\ka\eta_a^2}{\A_a(1+\eta_a^2)},\quad a=L,R.\label{214c}\ee
 Note, in particular, that the positivity of the left-hand-sides of Eqs.\eqref{453}
 makes $\alpha$ to belong to an open interval $]0,1[$   for $\ka>1$.

 \smallskip
 
 \noindent 3) We show that the $\sigma$-model \eqref{210c}
 remains classically integrable if we emancipate the parameter $\al$, that is, if we no longer consider the parameter $\al$ as the function of the parameters $\kappa,\eta_L,\eta_R$.  In what follows, we shall call the $\sigma$-model \eqref{210c} the bi-Yang-Baxter deformation of the WZW model\footnote{The reason  for this terminology is the fact that in the case $\al=0$ we recover from \eqref{210c} the standard WZW model. In what follows, we shall say "the DHKM model" whenever the TsT matrix is switched on. However, for the case of the vanishing TsT matrix we reserve the terminology "the bi-YB-WZ model".}  if the parameter $\alpha$ is emancipated and it belongs to the   interval $]-1,1[$. The  bi-Yang-Baxter deformation of the WZW model thus depends on four free parameters:  the  positive integer  $\kappa$ and three real numbers $\alpha$, $\rho_L$, $\rho_R$ the {\it absolute values} of which take values respectively in the  intervals  $[0,1[$, $]0,\pi[$ and $]0,\pi[$. 
 
 \smallskip
 
\noindent  4) We introduce the parameter $\alpha$ and emancipate it in the way compatible with integrability also in the presence of a nontrivial TsT matrix $\omega\neq 0$.
 
 \smallskip
 
 \noindent 5) We prove the renormalizability of  the bi-YB-WZ model  \eqref{210c} by showing that the RG flow concerns just the parameter $\alpha$, while the parameters $\ka$, $\rho_L$ and $\rho_R$ are renormalization group invariants.  We find the flow of $\al$ explicitly for every group target $K$ and show  that    for the special case of  $K=SU(2)$ the obtained  flow coincides with the RG flow  of the Lukyanov model described in Ref.\cite{L12}.
 
 \smallskip
 
\noindent  6) All the  results mentioned above are obtained by using the formalism of the so called $\E$-models \cite{KS96a,K15} as well as of their degenerate variants called the dressing cosets \cite{KS96b}.  In this paper, we introduce a new  method how to obtain the degenerate $\E$-models   out from the non-degenerate ones and we apply this method to prove that the bi-YB-WZ model is in fact an appropriate dressing coset. It is the latter circumstance which makes possible to prove   its renormalizability effortlessly.

\section{Introduction}
\setcounter{equation}{0}

Integrable deformations of nonlinear $\sigma$-models on group manifolds and on coset spaces constitute presently a topic
of  intense research activity. The subject originated long time ago in Refs.\cite{C81,Mad93,Fa}, where several deformations of the principal chiral model on the $SU(2)$ target were constructed, and several other $SU(2)$ results were subsequently obtained in Refs. \cite{FOZ,Moh,L12,KY,KMY11}. The study of  the integrability of $\sigma$-models living on higher dimensional group targets was initiated in Refs. \cite{K} by the present author, where  we introduced the so called $\eta$-deformations, induced in an appropriate way by solutions of the (modified) Yang-Baxter equation on the Lie algebra of the target group. This $\eta$-deformation algorithm, combined also with the coset construction of Refs.\cite{DMV13}  and with the alternative $\lambda$-deformation one \cite{S14},  gave rise to various  constructions of the  deformed integrable $\sigma$-models \cite{DST,KMY11,DMV15,DHKM,SST15,SSi,GS,GSS,DLMV,By,V}, many of them exploitable in quantum field theory \cite{ABR,DDST,LS,Lit,Fa19,FaLi,BKL,AHPT,ABF} and in string theory via the AdS/CFT correspondence \cite{AOSSY,BM,BRT,CL,GPZ,H,Ha,DMV13,DZ,LO,LT, NR,VT,SUY,DHT,SS,TSSY,ORSY}.

It turns out that the  Hamiltonian dynamics of many
 integrable  $\sigma$-models  can be  cast in a very transparent way  within the formalism of certain specific first order dynamical systems referred to as the $\E$-models \cite{KS96a,K15}. The $\E$-models are formulated in terms of the current algebras of Drinfeld doubles and  were originally introduced in the framework of the Poisson-Lie T-duality \cite{KS95,KS96a}. However,
they  turn out to be useful in many respects also in the integrability story, in particular in establishing the relation between the $\eta$ and $\lambda$ deformations via the T-duality
\cite{HT,SST15,K15}. 

Recently, Delduc, Hoare, Kameyama and Magro have found the multi-parametric integrable $\sigma$-model \eqref{438} living on an arbitrary simple group manifold $K$ \cite{DHKM}. In their approach,   they succeeded to merge consistently  several deformation procedures studied previously in a separate way, like the (bi)-Yang-Baxter deformations \cite{K},
the addition of the WZW term \cite{DMV15} or  the introduction of  the so-called TsT matrix \cite{F,LM,MY,VT,OV}. For the special case of the group
$SU(2)$, their result fits into the framework of the Lukyanov model \cite{L12}.

We show in Section 4 of the present paper that there exists an $\E$-model description of
the  DHKM $\sigma$-model (for the emancipated parameter $\alpha$), however, there is a novel element in the game comparing  with the cases of the low number of  deformation parameters treated in \cite{K15,K17}. Namely, the  $\E$-model underlying the DHKM $\sigma$-model turns out to be {\it degenerate}, that is, it is  the so called dressing coset  in the sense of Ref.\cite{KS96b}. 

Actually, we introduce in the present work  a new method of constructing the dressing cosets which is based on an appropriate  isotropic gauging of the non-degenerate $\E$-models.
This new approach is  technically very friendly and it plays the key role in the understanding of the structure of the DHKM model. We describe it in Section 3.3, just after reviewing the theory of the non-degenerate 
$\E$-models in Section 3.1 as well as  the old theory of the dressing cosets in Section 3.2.

What is it good for to know that the first order Hamiltonian dynamics of a nonlinear  $\sigma$-model can be described in terms of a particular (degenerate) $\E$-model? Well, the immediate benefit of this knowledge is the fact that
the $\sigma$-model underlied by the $\E$-model is {\it automatically  renormalizable} \cite{VKS,SfS,SSD10}.  This  means, in particular,  that the ultraviolet corrections just let  flow the parameters of the model without spoiling the form of the Lagrangian. Moreover, the $\E$-model formalism permits to determine the renormalization group flow by a   simple method introduced in Ref. \cite{SSD10,SV}. Actually, we employ this method in Section 6 to establish the renormalizability of the  bi-YB-WZ model, after proving in Section 5 its integrability.   Finally, we devote Section 6.4 to a detailed analysis of the case of $K=SU(2)$ where our results for the bi-YB-WZ RG flow are shown to match those of Lukyanov \cite{L12}.

 
\section{Dressing cosets}

\setcounter{equation}{0}

The dressing cosets construction \cite{KS96b} is the generalisation of the standard Poisson-Lie T-duality \cite{KS95} and  it was originally invented to produce new T-dual pairs of $\sigma$-models.  
While within the framework of the standard Poisson-Lie T-duality, the  Hamiltonian dynamics common to the mutually dual $\sigma$-models is that of an appropriate  $\E$-model \cite{KS96a,K15}, in the dressing cosets case, the Hamiltonian dynamics is that of a degenerate $\hat\E$-model in the sense of Ref. 
\cite{KS96b,K18}.  Although our concern in the present work is to deal with the degenerate $\hat\E$-models, we review also the non-degenerate case for reasons which are not merely pedagogical.
In fact, in Section 3.3 we introduce a new  method how to obtain the degenerate $\hat\E$-models (i.e. the dressing cosets) out from the non-degenerate ones. This new method is rapid and efficient and it lies at the basis of the understanding of the integrability and the renormalizability of the bi-YB-WZ model.

\subsection{Non-degenerate $\E$-models}
Consider a Lie group $D$ of even dimension $2d$ which is equipped with a bi-invariant Lorentzian metric of the signature $(d,d)$. This metric naturally induces a non-degenerate symmetric ad-invariant  bilinear form $(.,.)_\D$ on the Lie algebra $\D$ of $D$. A $d$-dimensional subgroup $K\subset D$ is called maximally isotropic if the restriction of the form $(.,.)_\D$ onto its Lie algebra $\K$ identically vanishes. If $D$  possesses a maximally isotropic subgroup $K$, the couple $(D,K)$  is called a Manin pair. If it possesses two (or more) maximally isotropic subgroups $K$, $\tilde K$ which are not connected by an internal automorphism of $D$, then $D$ is called the Drinfeld double.

We now associate certain infinite-dimensional symplectic manifold
$LD$ to every Drinfeld double $D$. The points of $LD$  are loops in $D$, that is maps $l:S^1\to D$ from a circle parametrized by the angle  variable $\sigma$ into the Drinfeld double $D$. For this reason, $LD$ is also known as the loop group of the Drinfeld double and it has itself the group structure given by the pointwise multiplication of the loops in $D$. It makes therefore sense to speak about the left-invariant   Maurer-Cartan form $l^{-1}dl$ on the group $LD$ and we can  define the symplectic form $\omega_{LD}$ on $LD$ by the formula
\be \omega_{LD}:=-\jp \oint  (l^{-1}dl,\partial_\sigma(l^{-1}dl))_\D.\label{173}\ee

The (non-degenerate) $\E$-model is a dynamical system the phase space of which is the symplectic manifold $(LD,\omega_{LD})$ and the Hamiltonian $H_\E$ of which is given by the formula 
\be H_\E=\jp\oint  (\partial_\sigma ll^{-1},\E \partial_\sigma ll^{-1})_\D.\label{176}\ee
Here $\E:\D\to\D$ is a $\br$-linear operator on the Lie algebra $\D$ of the double $D$. It has three important properties : 1) it squares to the identity operator on $\D$, i.e. $\E^2=$ Id; 2) it is self-adjoint with respect to the bilinear form $(.,.)_\D$, i.e. $(\E x,y)_\D=(x,\E y)_\D$, $x,y\in \D$; 3) the $\E$-dependent symmetric  bilinear form on $\D$ defined as   $(.,\E .)_\D$ is strictly positive definite. 

The knowledge of the symplectic form \eqref{173} and of the Hamiltonian \eqref{176} is sufficient to construct the first-order action of the $\E$-model \cite{KS96a}
\be S_\E(l)=\jp\int d\tau\oint (\partial_\tau ll^{-1},\partial_\sigma ll^{-1})_\D+\frac{1}{4}\int d^{-1}\oint (dll^{-1}\stackrel{\wedge}{,}[\partial_\sigma ll^{-1}, dll^{-1}])_\D-\jp\int d\tau\oint (\partial_\sigma ll^{-1},\E\partial_\sigma ll^{-1})_\D.\label{186} \ee
We note the presence of the WZ term in the action. Depending on the choice of the bilinear form $(.,.)_\D$, this term  may require a discrete overall normalisation in order to define a consistent quantum theory. We shall have more to say about this issue in Section 4. 

Every $\E$-model $(LD,\omega_{LD},H_\E)$ on the Drinfeld double represents simultaneously   the Hamiltonian dynamics of two (or more) $\sigma$-models living on geometrically non-equivalent targets. How it comes about? We show  this first  in a particular case of the so-called {\it perfect} Drinfeld doubles. Recall that the Drinfeld double $D$ is perfect if  the topological direct product $K\times\tilde K$ of its maximally isotropic subgroups is   diffeomorphic to $D$ in a way compatible with the multiplication law in $D$. This means that if $\Upsilon:D\to K\times\tilde K$ is the diffeomorphism and $m:D\times D\to D$ is the group multiplication map then the composition map $m\circ \Upsilon$ is the identity map on $D$. In particular, every element $l(\sigma)$ of the loop group $LD$ of the perfect Drinfeld double $D$ can be unambiguously decomposed as the product of one element $k(\sigma)$ from the loop group $LK$ and one element $\tilde h(\sigma)$ from the loop group $L\tilde K$ as follows
 \be l(\sigma)=k(\sigma)\tilde h(\sigma), \qquad k\in LK, \quad \tilde h\in L\tilde K.\label{dd}\ee 
 Inserting the decomposition \eqref{dd} into \eqref{173} and into \eqref{176}, we obtain easily
 \be \omega_{LD}=d\left(\oint (\partial_\sigma \tilde h\tilde h^{-1},k^{-1}dk)_\D\right),\label{sf1}\ee
\be H_\E(k,\tilde h) =\jp \oint (\partial_\sigma kk^{-1}+ k\partial_\sigma \tilde h\tilde h^{-1} k^{-1},\E(\partial_\sigma kk^{-1}+ k\partial_\sigma \tilde h\tilde h^{-1} k^{-1}))_\D.\label{hpm}\ee

The first order action \eqref{186} of the $\E$-model $(LD,\omega_{LD},H_\E)$ in the parametrization $k,\tilde h$ is therefore given by  the data
\eqref{sf1} and \eqref{hpm}:
\be S_\E=\int d\tau \oint   (\partial_\sigma \tilde h\tilde h^{-1},k^{-1}\partial_\tau k)_\D- \int d\tau H_\E(k, \tilde h ). \label{foa}\ee
The dependence of $S_\E$ on $\partial_\sigma \tilde h\tilde h^{-1}$ is quadratic, it is therefore easy to eliminate $\partial_\sigma \tilde h\tilde h^{-1}$  which gives the second order action  of the so called Poisson-Lie $\sigma$-model: 
 \be S_E(k)=\jp\int d\tau \oint  
 \left(\left(E+\Pi(k)\right)^{-1}\partial_+kk^{-1}, \partial_-kk^{-1}\right)_\D.\label{ea}\ee
 Here $\partial_\pm\equiv\partial_\tau\pm\partial_\sigma$,
  the linear operator $E:\tilde\K\to\K$ is such that its graph $\{\tilde x+E\tilde x,\tilde x\in\tilde\K\}$ coincides with the image of the operator Id$+\E$  and  
the $k$-dependent operator $\Pi(k):\tilde\K\to\K$ can be explicitly expressed in terms of the structure of the Drinfeld double as follows
\be \Pi(k)=-\J{\rm Ad}_k\tilde\J {\rm Ad}_{k^{-1}}\tilde\J.\label{202}\ee
Here Ad$_k$ stands for the adjoint action on $\D$ of the element $k\in K\subset D$ and   $\J,\tilde\J$ are projectors; $\J$ projects to $\K$ with the kernel $\tilde\K$ and  $\tilde\J$ projects to $\tilde\K$ with the kernel $\K$.

Recall also that the operator $\Pi(k):\tilde\K\to\K$ encodes the so called Poisson-Lie bracket of two functions $f_1,f_2$  on the group $K$ in the sense of the formula:
\be \{f_1,f_2\}_{K}(k)=(\nabla^L f_1,\Pi(k)\nabla^L f_2)_\D.\label{spls'}\ee
Here $\nabla^{L}$ is $\tilde\K$-valued differential operator acting on the functions on $K$  as
\be (\nabla^L f, x)_\D(k):= (\nabla^L_{x}f)(k)\equiv \frac{df(e^{sx}k)}{ds}\bigg\vert_{s=0}, \qquad x\in\K.\label{431}\ee

 Of course, every element $l(\sigma)$ of the loop group $LD$ can be decomposed also in the dual way as
 \be l(\sigma)=\tilde k(\sigma)h(\sigma), \qquad \tilde k\in L\tilde K, \quad h\in LK.\label{208}\ee 
 Inserting the decomposition \eqref{208} into \eqref{173} and into \eqref{176}, and then  repeating  all the procedure as before leads to the dual $\sigma$-model living on the target $\tilde K$:
  \be \tilde S_{\tilde E}(\tilde k)=\jp\int d\tau \oint  
 \left(\left(\tilde E+\tilde\Pi(\tilde k)\right)^{-1}\partial_+\tilde k\tilde k^{-1}, \partial_-\tilde k\tilde k^{-1}\right)_\D,\label{211}\ee
 where
 \be \tilde\Pi(\tilde k)=-\tilde\J{\rm Ad}_{\tilde k}\J {\rm Ad}_{\tilde k^{-1}}\J.\label{213}\ee
 Of course, the linear operator $\tilde E:\K\to\tilde \K$ is again such that its graph $\{ x+\tilde E x,  x\in\K\}$ coincides with the image of the operator Id$+\E$, which implies that the duality between the models \eqref{ea} and \eqref{211} holds under the condition, that the operator $\tilde E$ is inverse of the operator $E$.
 
 If the Drinfeld double is not perfect, there exists a generalization of the   T-duality between the models \eqref{ea} and \eqref{211}, where the two $\sigma$-models live, respectively, on the spaces of cosets $D/\tilde K$ and
 $D/K$ \cite{KS97,K15}. If we parametrize (possibly patch by patch) the coset space 
 $D/\tilde K$ by a section $m\in D$ of the bundle $D\to D/\tilde K$ and the coset space 
 $D/K$ by a section $\tilde m\in D$ of the bundle $D\to D/K$,
 then  the decompositions $l=m\tilde h$ and $l=\tilde mh$ generalize those \eqref{dd} and \eqref{208} and lead respectively to the following dual pair of $\sigma$-models with the WZW terms:
  \be S_\E(m)=\frac{1}{4} \int d\tau\oint \biggl(\Bigl(1-2P_m(\E)\Bigr)m^{-1}\partial_+ m,m^{-1}\partial_- m\biggr)_\D +\frac{1}{4}\int d^{-1}\oint \biggl(m^{-1}dm,[m^{-1}\partial_\sigma m,m^{-1}dm]\biggr)_\D;\label{219}\ee
    \be \tilde S_\E(\tilde m)=\frac{1}{4} \int d\tau\oint \biggl(\Bigl(1-2\tilde P_{\tilde m}(\E)\Bigr)\tilde m^{-1}\partial_+ \tilde m,\tilde m^{-1}\partial_- \tilde m\biggr)_\D +\frac{1}{4}\int d^{-1}\oint \biggl(\tilde m^{-1}d\tilde m,[\tilde m^{-1}\partial_\sigma \tilde m,\tilde m^{-1}d\tilde m]\biggr)_\D.\label{220}\ee
 Here the projectors $P_m(\E):\D\to\D$ and $\tilde P_{\tilde m}(\E):\D\to\D$ have the respective images $\tilde\K$ and $\K$
 and their respective kernels are given by the linear spaces (Id+Ad$_{m^{-1}}\E$Ad$_{m})\tilde\K$ and (Id+Ad$_{\tilde m^{-1}}\E$Ad$_{\tilde m})\K$.

 Of course, the Poisson-Lie  T-duality relating the models \eqref{ea} and \eqref{211}, or, more generally, relating the models \eqref{219} and \eqref{220},  is the main result that we review in this Section 2.1, but we add few more formulas about the $\E$-models  that will be useful in what follows.
 First of all, the first order Hamiltonian equations of motions derived from the formulas \eqref{173} and  \eqref{176} can be written in two useful ways, either as
 \be \partial_\tau ll^{-1}=\E\partial_\sigma ll^{-1}\label{216}\ee
 or as
 \be \partial_\tau j=\partial_\sigma(\E j)+[\E j,j],\label{218}\ee
 where
 \be j(\sigma):= \partial_\sigma l(\sigma)l(\sigma)^{-1}.\ee
 Moreover, the inversion of the symplectic form $\omega_{LD}$
 gives the standard current algebra Poisson brackets for the $L\D$-valued variable $j(\sigma)$:
 \be \{j^A(\sigma),j^B(\sigma')\}=F^{AB}_{~~C}j^C(\sigma)\delta(\sigma-\sigma')+D^{AB}\partial_\sigma\delta(\sigma-\sigma').\label{223}\ee
 Here
 \be j^A(\sigma):=(j(\sigma),T^A)_\D, \quad  [T^A,T^B]=F^{AB}_{~~C}T^C, \quad D^{AB}:=(T^A,T^B)_\D\ee
 and $T^A\in\D$ is some basis of $\D$.
 
 \subsection{Degenerate $\hat\E$-models}

We start the exposition of the degenerate $\hat\E$-models from the end, that is, we first write to what kind of T-duality they give rise to. To grasp the idea, it is sufficient to consider the perfect Drinfeld doubles and the resulting dual pair of the "dressing cosets" $\sigma$-models is then given by the actions
 \be S_{\hat E}(k)=\jp\int d\tau \oint   
 \left(\left({\rm Id}+\hat E\Pi(k)\right)^{-1}\hat E\partial_+kk^{-1}, \partial_-kk^{-1}\right)_\D,\label{244}\ee
  \be \tilde S_{{\hat E}}(\tilde k)=\jp\int d\tau \oint  
 \left(\left( {\hat E}+\tilde\Pi(\tilde k)\right)^{-1}\partial_+\tilde k\tilde k^{-1}, \partial_-\tilde k\tilde k^{-1}\right)_\D.\label{246}\ee
 Well, if $\hat E:\K\to\tilde\K$ is an invertible operator then  the dressing cosets models  \eqref{244} and \eqref{246} look identical to the models \eqref{ea} and \eqref{211}, so what is then new in the dressing coset story? Is it just the circumstance  that we release the condition that the operator ${\hat E}$ be invertible? No, this is not the case. In fact,  even if $\hat E$ is invertible, the dressing cosets pair \eqref{244} and \eqref{246} may contain substantially different physics as
 the standard Poisson-Lie T-dual pair \eqref{ea} and \eqref{211}. What happens is that 
in the construction  of the standard  pair  
\eqref{ea} and \eqref{211} from the $\E$-model it follows automatically that the symmetric part $\jp(\tilde E+\tilde E^*)$ of the operator  $\tilde E: \K\to\tilde \K$ is also invertible  (the symbol $\tilde E^*$ stands for the adjoint of the operator  $\tilde E$ with respect to the bilinear form $(.,.)_\D$.) In the dressing cosets case, the operator  $\jp(\hat E+\hat E^*)$ need not be invertible and the T-duality of the models  \eqref{244} and \eqref{246} still holds (if some further invariance conditions on $\hat E$ are fulfilled). This is not a trivial generalization of the standard Poisson-Lie T-duality, however, because the lack of invertibility of the operator $\hat E$ or of its symmetric part  $\jp(\hat E+\hat E^*)$  has drastic consequences on the dynamics of the mutually dual $\sigma$-models  \eqref{244} and \eqref{246}. In fact, both models develop a gauge symmetry with respect to some subgroup $F$ of the Drinfeld double $D$ and the common dimension of their targets thus gets effectively diminished by the dimension of the group $F$.

   Let us now give a concrete example of the phenomenon that the common gauge symmetry of the dressing cosets leads to the diminution of the  dimension of the targets of the mutually dual dressings cosets $\sigma$-models.  Consider thus the case where $K$ is a simple compact group  and  the group $\tilde K$ is the Lie algebra $\K$ with the (Abelian) group structure given by the vector space addition. As the manifold, the perfect Drinfeld double $D$ is the topological  direct product $K\times  \K$ with the following multiplication law
   \be (k_1,\kappa_1)(k_2,\kappa_2)=(k_1k_2,\kappa_1+{\rm Ad}_{k_1}\kappa_2), \qquad k_1,k_2\in K,\quad \kappa_1,\kappa_2\in\K.\ee
   Every element $k$ of the group $K$ is embedded in $D$ as $(k,0)$ and every element $\kappa$ of $\tilde K$ as $(e,\kappa)$, where $e$ is the unit element of $K$. The bilinear form $(.,.)_\D$  is given by
   \be \Bigl((\mu_1,\kappa_1),(\mu_2,\kappa_2)\Bigr)_\D=(\mu_1,\kappa_2)_\K+
   (\mu_2,\kappa_1)_\K,\quad \mu_{1,2},\kappa_{1,2}\in\K,\ee
   where $(.,.)_\K$ is the standard Killing-Cartan form on the simple Lie algebra $\K$.
   
   For the linear operator $ {\hat E}$, we pick the orthogonal projector $P^\perp$ on the subspace $\cal{T}^\perp\subset \K$ perpendicular to the Cartan subalgebra $\cal{T}\subset\K$, moreover, by using the formulas \eqref{202} and \eqref{213}, we find that  the Poisson-Lie bivector $\Pi(k)$ on $K$  trivially vanishes while the Poisson-Lie bivector on $\tilde K$  is given by the adjoint action of the Lie algebra. The actions of the mutually dual $\sigma$-models \eqref{244} and \eqref{246} in this particular case thus become 
\be S(k)=\jp\int d\tau\oint   \biggl( P^\perp\partial_+k k^{-1}, \partial_- k k^{-1}\biggr)_\K;\label{1131}\ee 
\be \tilde S(\kappa)=\jp\int d\tau\oint    \biggl( \Bigl(P^\perp- {\rm ad}_\kappa \Bigr)^{-1}\partial_+\kappa  , \partial_- \kappa \biggr)_\K.\label{1132}\ee 
The $\sigma$-models \eqref{1131} and \eqref{1132} have both the gauge symmetry with  the gauge group $F$ being the Cartan torus $\mathbb T\subset K$ (the Lie algebra of $F$ is the Cartan subalgebra $\cal{T}$). The element $f(\tau,\sigma)$ of the gauge group acts on the respective fields of the $\sigma$-models as
\be k\to fk, \qquad \kappa\to {\rm Ad}_f\kappa.\ee
In the  case of the group $K=SU(2)$, the common dimension of the targets  of the models  \eqref{1131} and \eqref{1132} is two and the corresponding background geometries were  obtained also in the framework of the standard non-Abelian T-duality \cite{OQ92,GR93,AABL93,Hew96} where the isometry group does not act freely. Therefore the dressing cosets in general can be understood as the Poisson-Lie generalizations of such models. Other examples of the dressing cosets have been studied  in \cite{S98,S99,KP99, CM06, HT, SST15, BTW, HS,SV}. 
 
 The first order dynamics of the dressing cosets was described in Ref.\cite{KS96b} and we now review that construction here. The phase space
 $LD_F$ of the  mutually dual pair of the dressing cosets is an appropriate symplectic reduction of the non-degenerate phase space $(LD,\omega_{LD})$. More precisely, consider a subgroup $F$ of $D$ which is isotropic, which means that the restriction of the bilinear form $(.,.)_\D$ to the Lie algebra $\F$ vanishes. The set of moment maps generating the left action of the loop group $LF$ on the loop group $LD$ is expressed by the quantity 
 $(\partial_\sigma ll^{-1},\F)_\D$. We set this quantity to $0$ which gives the presymplectic submanifold denoted by  $LD_F$. Said in other words,
the (pre)phase space $LD_F$  of the degenerate $\hat\E$-model is the space of the elements $l(\sigma)$ of the loop group $LD$, for which it holds for every $\sigma$
\be \d_\sigma ll^{-1} \in\F^\perp.\ee
Here the orthogonality symbol $\perp$ is understood with respect to the non-degenerate bilinear form $(.,.)_\D$.

The (pre)symplectic form of the degenerate $\hat\E$-model is just the restriction to  $LD_F$ of the symplectic form 
\be \omega_{LD}=-\jp \oint    (l^{-1}dl,\partial_\sigma(l^{-1}dl))_\D\label{1158}\ee
The Hamiltonian of the degenerate $\hat\E$-model looks the same as in the case of the standard Poisson-Lie T-duality  
\be H_{\hat\E}=\jp\oint   (\partial_\sigma ll^{-1},\hat\E \partial_\sigma ll^{-1})_\D,\label{1160}\ee
 however, the linear operator $\hat\E$ has now different properties as its counterpart $\E$ in the non-degenerate case. Here are those properties: 
$\hat\E:\F^\perp\to\F^\perp$ must be self-adjoint, it must commute with the adjoint action of  $\F$ on $\F^\perp$ and its kernel must  contain
$\F$.
Moreover,  the bilinear form $(.,\hat\E.)_{\F^\perp}$ must be positive semi-definitive and the image of the operator $\hat\E^2-{\rm Id}$ has to be contained in $\F$.

If the double is perfect, we can decompose the elements $l(\sigma)\in LD_F$ either as $l=k\tilde h$ or as  $l=\tilde kh$ and by eliminating respectively the
fields $\tilde h$ and $h$, we obtain the $\sigma$-models \eqref{244} and \eqref{246}. The linear operator $\hat E$ is  obtained via the relation 
\be {\rm Im}(\hat\E+\hat\E^2)\oplus\F =\{  x + {\hat  E} x,\  x\in\K\}.\label{296}\ee
After the fixing the gauge symmetry, the target spaces of the models \eqref{244} and \eqref{246} become respectively the dressing cosets $F\backslash K$ and
$F\backslash \tilde K$. The left dressing action of an element $f\in F$ on an element $k\in K$ is defined  as the $K$-part of the $D$-product $fk$.  Said in other words, we decompose $fk\in D$ as $^fk ^f\tilde h$, $^f k\in K$, $^f\tilde h\in \tilde K$ and the element
$^fk\in K$ is the result of the dressing action of $f$ on $k$.

How all  this procedure works in detail can be found in the original paper \cite{KS96b}, but the reader need not consult it. In fact, we shall present in the next Section 3.3 a new construction of the dressing cosets, which is arguably more   straightforward than that of Ref.\cite{KS96b}.
We do not know whether the new method permits to derive all dual pairs obtainable by the old one,  the new picture is however sufficiently general to underlie the bi-YB-WZ model \eqref{210c}.
\subsection{New method of producing the dressing cosets}
One particular way how to obtain the degenerate $\hat\E$-model from non-degenerate one was studied in \cite{S98}. The idea described therein is to let a non-degenerate operator $\E$ depend on a parameter and study a (singular) limit in which $\E$ becomes an operator $\hat\E$ characterising the degenerate  $\hat\E$-model. Here we develop another method, inspired by the procedure of the isotropic gauging described in Ref.\cite{KT94}, that is, we gauge in an isotropic way   a non-degenerate $\E$-model  to  produce  from it  a degenerate $\hat\E$-model.
The big advantage of this procedure is its simplicity as well as the  rapidity with which the resulting pair of the dressing cosets $\sigma$-models is explicitely found.  Let us show how it works.

Let $(LD,\omega_{LD},H_{\E})$ be a non-degenerate $\E$-model. Its first order action can be written as in \eqref{186}
\be S_\E(l)=\jp\int d\tau\oint (\partial_\tau ll^{-1},\partial_\sigma ll^{-1})_\D+\frac{1}{4}\int d^{-1}\oint (dll^{-1}\stackrel{\wedge}{,}[\partial_\sigma ll^{-1}, dll^{-1}])_\D-\jp\int d\tau\oint (\partial_\sigma ll^{-1},\E\partial_\sigma ll^{-1})_\D.\label{307} \ee
Let $F$ be an isotropic subgroup of $D$ (the term "isotropic" means that the restriction of the bilinear form $(.,.)_\D$ to the Lie algebra $\F$ vanishes) and let $\E$ be such that it commutes with the adjoint action of $F$ on $\D$.  The action \eqref{307} has then a global $F$-symmetry, where an element $f\in F$ acts on the loop $l(\sigma)$ by the standard left multiplication $fl(\sigma)$.  This global $F$-symmetry can be gauged\footnote{Due to the fact that $F$ is the isotropic subgroup the gauging is non-anomalous as it was thoroughly explained in Ref.\cite{KT94}.} by introducing an $\F$-valued gauge field $A\equiv A_\tau d\tau+A_\sigma d\sigma$ The gauged action reads
$$ S_\E(l,A)=\jp\int d\tau\oint (\partial_\tau ll^{-1}-2A_\tau,\partial_\sigma ll^{-1})_\D+$$\be +\frac{1}{4}\int d^{-1}\oint (dll^{-1}\stackrel{\wedge}{,}[\partial_\sigma ll^{-1}, dll^{-1}])_\D-\jp\int d\tau\oint (\partial_\sigma ll^{-1}-A_\sigma,\E(\partial_\sigma ll^{-1}-A_\sigma))_\D.\label{309} \ee
It is gauge invariant with respect to the following action of the element $f(\sigma,\tau)$ of the gauge group $F$ 
\be l\to fl,\quad A\to fAf^{-1}+dff^{-1}.\ee
To verify it, the $F$-invariance of the operator $\E$ is needed as well as the Polyakov-Wiegmann formula
\be W(fl)=W(l)+2d\oint \left((f^{-1}df,\partial_\sigma ll^{-1})_\D
-(dll^{-1},f^{-1}\partial_\sigma f)_\D\right),
\ee
where the $2$-form $W(l)$ on $LD$ is defined by
\be W(l)\equiv \oint (dll^{-1}\stackrel{\wedge}{,}[\partial_\sigma ll^{-1}, dll^{-1}])_\D.\ee
The basic claim of the present section is the statement:

\medskip

 {\it The isotropically gauged non-degenerate $\E$-model \eqref{309} is  the degenerate $\hat\E$-model.}

\medskip

\noindent Of course, the statement is deliberately short in order to encapsulate the message in the briefest terms, we have therefore explain it in more detail and also to state one more technical condition needed to be verified for the statement to hold. This condition is actually that the restriction of the non-degenerate  bilinear form $(.,\E.)_\D$ to the subalgebra $\F$ remains non-degenerate. 

It is not difficult to see  that the gauged $\E$-model \eqref{309} is in fact a degenerate $\hat\E$-model in a disguise. Indeed, the component $A_\tau$ play the role of the Lagrange multiplier which restricts the phase space $LD$
of the non-degenerate $\E$-model to the (pre)phase space
$LD_F=\{l\in LD, \partial_\sigma ll^{-1}\in\F^\perp\}$ of the degenerate one. Furthermore,
the field $A_\sigma$ appears quadratically in the action \eqref{309}, it can be therefore easily integrated away yielding again a quadratic Hamiltonian of the form $\oint (\partial_\sigma ll^{-1},\hat\E(\partial_\sigma ll^{-1}))_\D$ for some operator $\hat\E$. We have to show then that this operator $\hat\E$ has all the properties to define the degenerate $\hat\E$-model as described in the previous Section 3.2. 

We   perform the elimination of the field $A_\sigma$ by a chain of shortcut arguments, avoiding any "hardline" computation.
We first remark that  if it existed 
a non-vanishing $x\in\F$ such that $x=\E y$ for some $y\in\F$,
then we would have $\E(x\pm y)= y\pm x$, and
the expressions $(x\pm y,\E(x\pm y))_D$ would both vanish which would contradict   the strict positive definiteness of the bilinear form $(.,\E.)_\D$. We infer that 
the linear space $\E\F$ has trivial (zero) intersection with $\F$ and we  can therefore set 
\be V=\F\oplus\E\F.\ee
Let us now argue that the vector spaces $V$ and $V^\perp$ also intersect trivially. Indeed, if the intersection $V\cap V^\perp$ 
contained a non-zero vector $x+\E y$, $x,y\in\F$ then every vector $z\in\F\subset V$  would be orthogonal to $x+\E y$, hence to $\E y$ which would be contradictory to the fact that the restriction of the bilinear form $(.,\E.)_\D$ to $\F$ be non-degenerate.
It follows that we can write the Lie algebra $\D$ as the direct sum 
$V^\perp \oplus \F \oplus \E\F $  and  represent every element $y\in\D$ accordingly as
\be y=y_0+y_1+y_2.\label{396b}\ee The action \eqref{309} can be now rewritten
as $$S_\E(l,A)=\jp\int d\tau\oint (\partial_\tau ll^{-1}-2A_\tau,\partial_\sigma ll^{-1})_\D+ \frac{1}{4}\int d^{-1}\oint (dll^{-1}\stackrel{\wedge}{,}[\partial_\sigma ll^{-1}, dll^{-1}])_\D$$\be -\jp\int d\tau\oint \Bigl((\partial_\sigma ll^{-1}) _0, \E(\partial_\sigma ll^{-1})_0\Bigr)_\D  -\jp\int d\tau\oint \Bigl(\Bigl((\partial_\sigma ll^{-1})_1 -A_\sigma\Bigr), \E\Bigl((\partial_\sigma ll^{-1})_1 -A_\sigma\Bigr)\Bigr)_\D.\label{345} \ee
Note that the component $(\partial_\sigma ll^{-1})_2$ does not appear in the Hamiltonian part of the action because it is killed by the Lagrange multiplier $A_\tau$. Furthermore, the component $(\partial_\sigma ll^{-1})_1$ lies in $\F$, it can be therefore absorbed into $A_\sigma$.  Integrating away $A_\sigma$ thus means simply the omitting of the last term in Eq.\eqref{345}.
At the end, we obtain  the degenerate $\hat\E$-model where
the operator $\hat\E:\F^\perp \to \F^\perp$ is defined as
\be \hat\E(y_0+y_1):=\E y_0.\label{349}\ee
It is easy to verify that $\hat\E$ has all required properties  to define the degenerate $\hat\E$-model. It is self-adjoint because $\E$ is:
\be (x_0+x_1,\hat\E(y_0+y_1))_\D=(x_0,\E y_0)_\D=
(\E x_0,y_0)_\D=  (\hat\E(x_0+x_1),y_0+y_1)_\D,\ee
its kernel evidently contains $\F$ and  it commutes with ad$_\phi$ for every $\phi\in\F$ because both $V$ and $V^\perp$ are ad$_\F$ invariant:
\be \hat\E[\phi,x_0+x_1]=\E[\phi,x_0]=[\phi,\E x_0]= [\phi,\hat\E(x_0+x_1)].\ee
Moreover, it holds
\be (\hat\E^2-{\rm Id})(x_0+x_1)=\E^2x_0 -x_0 -x_1=-x_1,\ee
therefore the image of the operator $(\hat\E^2-{\rm Id})$ is indeed contained in $\F$. Finally, the bilinear form $(.,\hat\E .)_{\F^\perp}$ is semi-positive definite because
\be (x_0+x_1,\hat\E(y_0+y_1))_\D=(x_0,\E y_0)_\D\ee
and the form $(.,\E.)_\D$ is strictly positive definite.

What is it good for to know that the gauged non-degenerate $\E$-model is in fact the degenerate $\hat\E$-model? Well, in some important cases, like those studied in the context of the integrable deformations, it is technically much easier to extract the actions of the  dual pair of $\sigma$-models  from the gauged non-degenerate first order formalism rather than directly from the degenerate one.
 We give now an example of this situation for the case when $D$ is the perfect double and $\F$ is a subalgebra of $\K$. 
 
 Instead of eliminating the gauge field $A$ from the gauged first order action \eqref{345}, we first decompose $l$
   as the product of one element $k(\sigma)$ of the loop group $LK$ and one element $\tilde h(\sigma)$ of the loop group $L\tilde K$ as in Eq.\eqref{dd}
 \be l(\sigma)=k(\sigma)\tilde h(\sigma), \qquad k\in LK, \quad \tilde h\in L\tilde K.\label{373}\ee 
 Inserting the decomposition \eqref{373} into \eqref{309}, we obtain easily
$$ S_\E(k,\tilde h, A)=\int d\tau \oint   (\partial_\sigma \tilde h\tilde h^{-1},k^{-1}(\partial_\tau k k^{-1} - A_\tau) k)_\D$$\be - \jp\int d\tau \oint (\partial_\sigma kk^{-1}-A_\sigma+ k\partial_\sigma \tilde h\tilde h^{-1} k^{-1},\E(\partial_\sigma kk^{-1}-A_\sigma+ k\partial_\sigma \tilde h\tilde h^{-1} k^{-1}))_\D.\label{375}\ee
It is easy to preform the computation integrating away  $\partial_\sigma \tilde h\tilde h^{-1}$ from $S_\E(k,\tilde h, A)$, because the only difference with respect to the similar computation leading from \eqref{foa} to \eqref{ea} is the
replacement of $\partial_\tau k k^{-1}$ by $\partial_\tau k k^{-1} - A_\tau$ and of $\partial_\sigma k k^{-1}$ by $\partial_\sigma k k^{-1} - A_\sigma$. The result is therefore the following gauged second order action  of the standard Poisson-Lie $\sigma$-model \eqref{ea}: 
 \be S_E(k,A)=\jp\int d\tau \oint  
 \left(\left(E+\Pi(k)\right)^{-1}(\partial_+kk^{-1}-A_+), \partial_-kk^{-1}-A_-\right)_\D.\label{378}\ee  
Here $A_\pm\equiv A_\tau\pm A_\sigma$  and, as before, the graph $\{\tilde x +E\tilde x, \tilde x\in\tilde\K\}$ of the operator $E:\tilde\K\to\K$ coincides with the image of the operator Id$+\E$.  The gauge symmetry $k\to fk$ and $A\to fAf^{-1}+dff^{-1}$ of the action \eqref{378} is evident.

Remarkably, the integrating away the non-dynamical gauge fields $A_\pm$ from \eqref{378} gives the dressing coset action \eqref{244}
 \be S_{\hat E}(k)=\jp\int d\tau \oint  
 \left(\left({\rm Id}+\hat E\Pi(k)\right)^{-1}\hat E\partial_+kk^{-1} , \partial_-kk^{-1} \right)_\D.\label{379}\ee  
We now perform in detail the calculation leading from \eqref{378} to \eqref{379} which will permit us to identify the operator $\hat E:\K\to\tilde\K$ in terms of the operator $E:\tilde\K\to\K$. We start by introducing an $\tilde\K$-valued auxiliary $1$-form field $B\equiv B_\tau d\tau + B_\sigma d\sigma$ and by considering an auxiliary action
\be S_E(k,A,B)=\jp\int d\tau \oint  
 \left(\left( E+\Pi(k)\right)B_+,B_-\right)_\D+ \int\left((dkk^{-1}-A)\stackrel{\wedge}{,}B\right)_\D.\label{382}\ee
 The auxiliary action $S_E(k,A,B)$ is dynamically equivalent to the action $S_E(k,A)$ because the integrating away the field $B$ from the former yields the latter. Now the gauge field $A$ featuring in the action \eqref{382} plays the role of the Lagrange multiplier making to vanish some components of the field $B$. Let us be more precise about that point:
 
 Let $\tilde Q:\tilde\K\to\tilde\K$ be the projector with the
 kernel $\E\F\cap\tilde\K$ and the image $(\F\oplus V^\perp)\cap\tilde\K$ and
 let   $Q:\K\to\K$ be the projector  with the kernel $\F\cap \K$ and the image $(\E\F\oplus V^\perp)\cap\K$. Integrating away the gauge field $A$ from the auxiliary action \eqref{382} then gives
 \be S_E(k,B)=\jp\int d\tau \oint  
 \left(Q\left( E+\Pi(k)\right)\tilde QB_+,\tilde QB_-\right)_\D+
 \int\left( dkk^{-1}\stackrel{\wedge}{,}\tilde Q B\right)_\D.\label{390}\ee
 Finally, integrating away $\tilde Q B$ from \eqref{390} yields the dressing coset action \eqref{379} where
 \be \hat E=(QE\tilde Q)^{-1}Q.\label{392}\ee
 Let us now recover the dual dressing coset \eqref{246} from the gauged $\E$-model \eqref{309}.
 We decompose $l$
   as the product of one element $\tilde k(\sigma)$ of the loop group $L\tilde K$ and one element $h(\sigma)$ of the loop group $LK$ as in Eq.\eqref{208}
 \be l(\sigma)=\tilde k(\sigma) h(\sigma), \qquad \tilde k\in L\tilde K, \quad   h\in LK.\label{396}\ee 
 Inserting the decomposition \eqref{396} into \eqref{309}, we obtain
$$ \tilde S_\E(\tilde k,h, A)=\int d\tau \oint   (\partial_\sigma h  h^{-1},\tilde k^{-1}(\partial_\tau \tilde k \tilde k^{-1} - A_\tau) \tilde k)_\D-\int d\tau\oint(A_\tau,\partial_\sigma\tilde k\tilde k^{-1})_\D $$\be - \jp\int d\tau \oint (\partial_\sigma \tilde k\tilde k^{-1}-A_\sigma+ \tilde k\partial_\sigma h h^{-1} \tilde k^{-1},\E(\partial_\sigma \tilde k\tilde k^{-1}-A_\sigma+ \tilde k\partial_\sigma  hh^{-1} \tilde k^{-1}))_\D.\label{398}\ee
We write $A_\sigma$ as
\be A_\sigma ={\rm Ad}_{\tilde k}\J {\rm Ad}_{\tilde k^{-1}}A_\sigma +\tilde\Pi(\tilde k)A_\sigma,\label{400}\ee
and introduce a $\K$-valued field $\Lambda$
\be \Lambda:= \partial_\sigma hh^{-1}-\J{\rm Ad}_{\tilde k^{-1}}A_\sigma.\label{402}\ee
We can now rewrite \eqref{398} as
$$ \tilde S_\E(\tilde k,\Lambda, A)=\int d\tau \oint   (\Lambda,k^{-1}(\partial_\tau \tilde k \tilde k^{-1} - \tilde\Pi(\tilde k)A_\tau) \tilde k)_\D-\int (A\stackrel{\wedge}{,} d\tilde k\tilde k^{-1})_\D +\jp\int (A\stackrel{\wedge}{,}\tilde\Pi(\tilde k)A)_\D $$\be - \jp\int d\tau \oint (\partial_\sigma \tilde k\tilde k^{-1}-\tilde\Pi(\tilde k)A_\sigma+ \tilde k\Lambda\tilde k^{-1},\E(\partial_\sigma \tilde k\tilde k^{-1}-\tilde\Pi(\tilde k)A_\sigma+ \tilde k\Lambda \tilde k^{-1}))_\D.\label{404}\ee
It is easy to perform the computation integrating away  $\Lambda$ from $\tilde S_\E(\tilde k,\Lambda, A)$, because the only difference with respect to the similar computation leading from \eqref{foa} to \eqref{211} is the
replacement of $\partial_\tau \tilde k \tilde k^{-1}$ by $\partial_\tau \tilde k \tilde k^{-1} - \tilde\Pi(\tilde k)A_\tau$, of $\partial_\sigma k k^{-1}$ by $\partial_\sigma k k^{-1} - \tilde\Pi(\tilde k)A_\sigma$ and of $\partial_\sigma hh^{-1}$ by $\Lambda$. The result is therefore the following gauged second order action  of the dual Poisson-Lie $\sigma$-model \eqref{211}: 
 $$\tilde S_E(\tilde k,A)=\jp\int d\tau \oint  
 \left(\left(E^{-1}+\tilde\Pi(\tilde k)\right)^{-1}(\partial_+\tilde k\tilde k^{-1}-\tilde\Pi(\tilde k)A_+), \partial_-\tilde k\tilde k^{-1}-\tilde\Pi(\tilde k)A_-\right)_\D+$$\be 
 +\jp\int (A\stackrel{\wedge}{,}\tilde\Pi(\tilde k)A)_\D -\int (A\stackrel{\wedge}{,} d\tilde k\tilde k^{-1})_\D \label{408}\ee  
Integrating away the gauge field $A$ gives
\be \tilde S_{{\hat E}}(\tilde k)=\jp\int d\tau \oint  
 \left(\left( {\hat E}+\tilde\Pi(\tilde k)\right)^{-1}\partial_+\tilde k\tilde k^{-1}, \partial_-\tilde k\tilde k^{-1}\right)_\D,\label{412}\ee
 where the operator $\hat E$ is given by Eq.\eqref{392}.
 
We summarize: starting from the gauged non-degenerate $\E$-model \eqref{309}, we have produced the dressing coset pair \eqref{379} and \eqref{412}, or equivalently, the pair \eqref{244} and \eqref{246}.
For completeness, let us check that the operator $\hat E$ given by \eqref{392} indeed verifies the relation \eqref{296}, if the operator
$\hat\E$ is given by Eq.\eqref{349}.  First of all we find that
\be {\rm Im} (\hat\E+\hat\E^2)\oplus \F= (1+\E)V^\perp\oplus \F\ee
We have to show that
\be \{x+(QE\tilde Q)^{-1}Qx, x\in \K\}=(1+\E)V^\perp\oplus \F,\ee
or, equivalently, to show that 
\be \{Qx+(QE\tilde Q)^{-1}Qx, x\in\K\}=(1+\E)V^\perp.\label{422}\ee
The right-hand-side of \eqref{422} is equal to $\{\tilde Q\tilde x+E\tilde Q\tilde x, x\in\tilde \K\}$,
we have to show therefore that
\be \{Qx+(QE\tilde Q)^{-1}Qx, x\in \K\}=\{\tilde Q\tilde x+E\tilde Q\tilde x, x\in\tilde \K\}.\ee
But this is evidently true since $E\tilde Q\tilde x\in V^\perp\cap \K$,
hence 
  $E\tilde Q\tilde x=(QE\tilde Q)\tilde x$.

  \section{DHKM model as the degenerate $\hat\E$-model} 
  \setcounter{equation}{0}
 
 The main concern of the present section is to show that the DHKM  $\sigma$-model \eqref{438} can be extracted from an appropriate   isotropically gauged   non-degenerate
 $\E$-model \eqref{309}  by the procedure  described in Section 3.3  (after Eq.\eqref{373}). Said in other words, for an appropriate  choice of the Drinfeld double $D$, of the non-degenerate operator $\E$, of the gauge group $F$ and of the maximally isotropic subgroup $\tilde K_\Omega\subset D$,  the  DHKM $\sigma$-model is the dressing coset \cite{KS96b}  living on the double coset target $F\backslash D/ \tilde K_\Omega$.  
 
We start with the description of the  relevant Drinfeld double $D$ which is the direct product $K^\bc\times K^\bc$, where $K^\bc$ denotes the complexification of the simple compact group $K$. The invariant bilinear form $(.,.)_\D$ on the Lie algebra $\D=\K^\bc\oplus \K^\bc$ is given by the formula
 \be \left(z_L\oplus z_R,z'_L \oplus z'_R\right)_\D:=\frac{4\ka}{\sin{(\rho_L)}}  \Im\tr\left(e^{\ri \rho_L}z_Lz'_L\right)+\frac{4\ka}{\sin{(\rho_R)}}\Im\tr\left(e^{-\ri \rho_R}z_Rz'_R\right),\quad z_a,z'_a\in \K^\bc, \ a=L,R.\label{498a}\ee
Here the symbol  $\Im$ means taking the imaginary  part of a complex number, $\ka$ is a positive integer and the absolute values of the real parameters $\rho_L,\rho_R$   range in the open interval $]0,\pi[$.  The positive integer choice of  $\ka$  
  guarantees the required  $2\pi$ ambiguity of the WZ term of the $\E$-model action which is needed for the consistent quantization. We note in this respect that
 the bilinear form \eqref{498a} can be  rewritten as 
\be\left(z_L\oplus z_R,z'_L \oplus z'_R\right)_\D:= 4\ka\cot{(\rho_L)}  \Im\tr\left( z_Lz'_L\right)+
 4\ka\cot{(\rho_R)}\Im\tr\left(z_Rz'_R\right)+4\ka\Re\tr\left( z_Lz'_L\right)-4\ka\Re\tr\left( z_Rz'_R\right). \label{589b}\ee
The  first part on the right-hand-side, containing $\Im\tr$, is of the Lu-Weinstein type and it does not require the discrete normalization because it does not lead to the presence of the WZ term at the level of the second order $\sigma$-model action. However, the $\Re\tr$ part does lead to the presence of the WZ term at the second order level and it must be therefore appropriately discretely normalized.
 
The next step is to specify the non-degenerate operator $\E:\D\to\D$. It is given by the formula
\be \E(z_L\oplus z_R)= \ri\left(\frac{1-\mu_L^2}{2\mu_L}z_L+e^{-\ri \rho_L}\frac{1+\mu_L^2}{2\mu_L}z_L^*\right)\oplus  \ri\left(\frac{1-\mu_R^2}{2\mu_R}z_R+e^{\ri \rho_R}\frac{1+\mu_R^2}{2\mu_R}z_R^*\right), \label{480a}\ee
where $z^*$ stands for the Hermitian conjugation and $\mu_L,\mu_R$ are real  parameters having, respectively, the same signs as the parameters
$\rho_L,\rho_R$.

We note that the operator $\E$ given by Eq. \eqref{480a}   is the direct some of two copies of the $\E$-operators used for the construction of the Yang-Baxter $\sigma$-model with the WZW term \cite{K17}.  It is straightforward to check that the operator \eqref{480a} verifies all three properties needed to define the non-degenerate $\E$-model, namely, it squares to  the identity,   it is self-adjoint with respect to the bilinear form \eqref{498a} and the 
bilinear form $(.,\E.)_D$ on $\D$ is strictly positive definite. 

The first order action $S_\E(l)$ of the non-degenerate $\E$-model defined by the data $(D,\E)$ is given by the general expression \eqref{186}. In order to recover the $\sigma$-model \eqref{438} out of it, we have to gauge it isotropically in the sense of Section 3.3, that is, to produce the gauge invariant first order  action $S_\E(l,A)$ given by  Eq. \eqref{309}.  For the   gauge group $F\subset D$, we choose the diagonal embedding of the simple compact group $K$ into $D=K^\bc\times K^\bc$. The elements of $F$ have therefore the form $(f,f)\in D$, $f\in K$ and the elements of the Lie algebra $\F$ have the form $x\oplus x\in \D$, $x\in\K$.  The gauge group $F$ is isotropic because  it is easy to check that it holds
\be (x\oplus x,y\oplus y)_\D=0,\quad \forall x,y\in\K.\ee
The operator $\E$ given by Eq.\eqref{480a} commutes with the adjoint action of the Lie group $F$ on $\D$ because it holds $(\Ad_f z)^*=\Ad_f (z^*)$ for $f\in K, z\in \K^\bc$. 
To fit into the general gauging procedure of Section 3.3, the last thing to check is that the restriction of the non-degenerate bilinear form $(.,\E.)_\D$ onto the subalgebra $\F$ remains non-degenerate. But this is true because this restriction is given by the formula\footnote{Note that the trace tr  on the compact Lie algebra is negative definite, moreover,  in order to obtain a consistent quantum theory, we normalize it in such a way that  the ambiguity in  the WZ term in \eqref{210c} is a multiple of $2\pi$.}  
\be (x\oplus x,\E (y\oplus y))_\F = - 2\ka\left(\mu_L\cot{\frac{\rho_L}{2}}+\mu_L^{-1}\tan{\frac{\rho_L}{2}}
+\mu_R\cot{\frac{\rho_R}{2}}+\mu_R^{-1}\tan{\frac{\rho_R}{2}}
  \right)\tr(xy).\ee

We  claim that the ingredients $D$, $\E$ and $F$ underlie the dressing coset coinciding with  the  DHKM $\sigma$-model \eqref{438} with the emancipated parameter $\alpha$. But if so, why $D$, $\E$ and $F$ do not depend on the TsT parameters?  It turns out that  the  $r^2$   TsT parameters   are not visible at the first order formalism but they appear in the process of the extraction of the  second order $\sigma$-model from the gauged action \eqref{309}. Speaking more precisely, it  is the choice of the maximally isotropic subgroup $\tilde K_\Omega$ of $D$ which depends on the TsT matrix $\Omega$ and this choice is needed to trigger the procedure starting by Eq. \eqref{373} and permitting to extract from the first order action
 \eqref{309} the second order $\sigma$-model living on the target $F\backslash D/\tilde K_\Omega$. Recall at the same time  that here we are touching the very core of the Poisson-Lie T-duality story: every (degenerate) $\E$-model gives rise to as many geometrically non-equivalent $\sigma$-models as is the number of maximally isotropic subgroups of the Drinfeld double $D$ which are not related by an inner automorphism. At the same time, all those geometrically inequivalent $\sigma$-models are dynamically equivalent  as the Hamiltonian systems  possibly up to the dynamics of a finite number of zero modes  determined by the string boundary conditions (for more details see Ref.\cite{KSopen}). In our particular case, a class of maximally maximally isotropic subgroups is
  parametrized by the TsT matrix 
 $\Omega:{\mathcal T}\to {\mathcal T}$.  Changing of the value of the matrix $\Omega$  from one to another is thus  the Poisson-Lie T-duality transformation; the  first order Hamiltonian dynamics of the $\sigma$-models remains independent of the choice of $\Omega$ (up to the finite number of degrees of freedom), however, their target space geometries do depend on the choice of $\Omega$. 
 
Now we describe in detail the maximally isotropic subgroup $\tilde K_\Omega\subset D$. First of all, it has  the form of the semidirect product $ A_\Omega \ltimes (N\times N)$, where
$N$ is the nilpotent subgroup of $K^\bc$ appearing in the Iwasawa decomposition $K^\bc=KAN$ and $A_\Omega$ is certain $\Omega$-dependent $r$-dimensional isotropic subgroup of the group $A^\bc \times A^\bc$.  Actually, the maximally isotropic subgroup $\tilde K_\Omega$ is fully determined by
its Lie algebra $\tilde\K_\Omega$ which can be conveniently described in terms of the Yang-Baxter operator $R$ as the following half-dimensional subspace of the double $\D=\K^\bc\oplus \K^\bc$
\be \tilde K_\Omega\!=\! \left\{e^{-\frac{\ri\rho_L}{2}}\left(\!(R-\ri)u_L+\frac{8\ka}{\sin{\rho_R}}\Omega^t(R^2+1)u_R\!\right)\!\oplus e^{\frac{\ri\rho_R}{2}}\left(\!(R-\ri)u_R-\frac{8\ka}{\sin{\rho_L}}\Omega(R^2+1)u_L\!\right);   u_L,u_R\in\K\right\},\label{493}\ee
Recall that $\Omega:\mathcal{T}\to\mathcal{T}$ is arbitrary and $\Omega^t$ stands for the transposition with respect to the bilinear form
on the Cartan subalgebra $\mathcal{T}$ defined by the trace. It is the matter of a simple check that the restriction of the bilinear form $(.,.)_\D$ on  $\tilde\K_\Omega$  vanishes.

It is perhaps interesting to make a small digression
and to represent the Lie algebra structure of $\tilde\K_\Omega$ as an alternative commutator $[.,.]_{R,\rho,\Omega}$ on the vector space $\K\oplus\K$
$$ [u_L\oplus u_R,v_L\oplus v_R]_{R,\rho,\Omega}=$$ $$=\left(\sin{\left(\frac{\rho_L}{2}\right)}R+\col\right)
\left([u_L,v_L]_R+\frac{8\ka}{\sin{\rho_R}}[\Omega^t(R^2+1)u_R,v_L]-\frac{8\ka}{\sin{\rho_R}}[\Omega^t(R^2+1)v_R,u_L]\right)\oplus$$
\be \oplus\left(-\sin{\left(\frac{\rho_R}{2}\right)}R+\cor\right)
\left([u_R,v_R]_R-\frac{8\ka}{\sin{\rho_L}}[\Omega(R^2+1)u_L,v_R]+\frac{8\ka}{\sin{\rho_L}}[\Omega(R^2+1)v_L,u_R]\right).\ee
Here the notation $[.,.]_R$ means
\be [u,v]_R:=[Ru,v]+[u,Rv].\label{505b}\ee
Note that for $\Omega=0$, the Lie algebra $\tilde\K_\Omega$
becomes the direct sum of two Lie algebras $\K_{R,\rho_L}\oplus \K_{R,\rho_R}$ characterized by the commutators 
\be [u,v]_{R,\rho_L}=\left(\sin{\left(\frac{\rho_L}{2}\right)}R+\col\right)
 [u,v]_R, \  [u,v]_{R,\rho_R}=\left(-\sir R+\cor\right)[u,v]_R, \  u,v\in\K.\ee

Consider now the first order action \eqref{309} of the isotropically gauged $\E$-model:
$$ S_\E(l,A)=\jp\int d\tau\oint (\partial_\tau ll^{-1}-2A_\tau,\partial_\sigma ll^{-1})_\D+$$\be +\frac{1}{4}\int d^{-1}\oint (dll^{-1}\stackrel{\wedge}{,}[\partial_\sigma ll^{-1}, dll^{-1}])_\D-\jp\int d\tau\oint (\partial_\sigma ll^{-1}-A_\sigma,\E(\partial_\sigma ll^{-1}-A_\sigma))_\D.\label{533a} \ee
Following the general procedure described in Section 3.3, we write the field $l\in K^\bc\times  K^\bc$ as
\be l=m\tilde h_\Omega,\label{500}\ee
where $m$ is $K\times K$-valued field and $\tilde h_\Omega$ takes values in $\tilde K_\Omega$. It is easy to see that the decomposition 
\eqref{500} is global for whatever $\Omega$. Inserting it into the action \eqref{533a}, we obtain\footnote{Note the presence of three more terms in the action \eqref{503} which are absent in Eq. \eqref{375}.  This is because in \eqref{375} we have considered  the field $k$ taking values in the maximally isotropic subgroup which is not  the case for our field $m$. }
$$ S_\E(m,\tilde h, A)=\jp\int d\tau\oint (\partial_\tau mm^{-1},\partial_\sigma mm^{-1})_\D+\frac{1}{4}\int d^{-1}\oint (dmm^{-1}\stackrel{\wedge}{,}[\partial_\sigma mm^{-1}, dmm^{-1}])_\D$$$$+\int d\tau \oint   (\partial_\sigma \tilde h_\Omega\tilde h_\Omega^{-1},m^{-1}(\partial_\tau m m^{-1} - A_\tau) m)_\D- \int dt\oint (A_\tau,\partial_\sigma mm^{-1})_\D+$$\be +\jp\int d\tau \oint (\partial_\sigma mm^{-1}-A_\sigma+ m\partial_\sigma \tilde h_\Omega\tilde h_\Omega^{-1} m^{-1},\E(\partial_\sigma mm^{-1}-A_\sigma+ m\partial_\sigma \tilde h_\Omega\tilde h_\Omega^{-1} m^{-1}))_\D.\label{503}\ee

We wish to integrate away the field $\partial_\sigma \tilde h_\Omega\tilde h_\Omega^{-1}$. In the case of
absence of the gauge field $A$, the result would be given by the general formula \eqref{219}. In the presence of $A$, the formula \eqref{219} has to be modified accordingly:
$$ S_\E(m,A))=+\frac{1}{4} \int d\tau\oint \biggl(\Bigl(1-2P_{\Omega,m}(\E)\Bigr)(m^{-1}\partial_+ m -m^{-1}A_+m),m^{-1}\partial_- m-m^{-1}A_-m\biggr)_\D+$$
\be + \frac{1}{4}\int d^{-1}\oint \biggl(m^{-1}dm,[m^{-1}\partial_\sigma m,m^{-1}dm]\biggr)_\D +\frac{1}{4} \int d\tau \oint (A_+,\partial_- mm^{-1})_\D-\frac{1}{4} \int d\tau \oint (A_-,\partial_+ mm^{-1})_\D, \label{509}\ee
where $P_{\Omega,m}(\E):\D\to\D$ is the projector with the image $\tilde\K_\Omega$ and the kernel
$({\rm Id}+\Ad_{m^{-1}}\E\Ad_m)\tilde\K_\Omega$.

It remains to determine explicitly the projector $P_{\Omega,m}(\E)$. We first note that there is no dependence of $m$ since the operator Ad$_m$ commutes with the
operator $\E$ defined by the formula \eqref{480a}; therefore $P_{\Omega,m}(\E)\equiv P_\Omega(\E)$. Then we find  that the kernel $({\rm Id}+\E)\tilde\K_\Omega$ is the half-dimensional subspace of $\D$
which can be parametrized by the elements $s_L\oplus s_R$ of the Lie algebra $\K\oplus\K$ via
$$({\rm Id}+\E)\tilde\K_\Omega=e^{-\ri\frac{\rho_L}{2}}(1+\ri\mu_L)s_L\oplus e^{\ri\frac{\rho_R}{2}}(1+\ri\mu_R)s_R.$$
We have to calculate the action of the projector $P_{\Omega}(\E)$ on the elements $x_L\oplus x_R$ of the Lie algebra $\K\oplus \K$; it is determined from the unambiguous decomposition 
$$x_L\oplus x_R= e^{-\ri\frac{\rho_L}{2}}(1-\ri\mu_L)s_L\oplus e^{\ri\frac{\rho_R}{2}}(1-\ri\mu_R)s_R + $$\be +
 e^{-\frac{\ri\rho_L}{2}}\left(\!(R-\ri)u_L+\frac{8\ka}{\sin{\rho_R}}\Omega^t(R^2+1)u_R\!\right)\!\oplus e^{\frac{\ri\rho_R}{2}}\left(\!(R-\ri)u_R-\frac{8\ka}{\sin{\rho_L}}\Omega(R^2+1)u_L\!\right).\label{559oa}\ee
 Said differently: 
  given $x_L\oplus x_R$, we have to find $s_L,s_R,u_L,u_R\in\K$ (they are given unambiguously) such that the relation \eqref{559oa} holds.
  Then we have
  \be P_\Omega(\E)(x_L\oplus x_R)=
 e^{-\frac{\ri\rho_L}{2}}\left(\!(R-\ri)u_L+\frac{8\ka}{\sin{\rho_R}}\Omega^t(R^2+1)u_R\!\right)\!\oplus e^{\frac{\ri\rho_R}{2}}\left(\!(R-\ri)u_R-\frac{8\ka}{\sin{\rho_L}}\Omega(R^2+1)u_L\!\right).\label{565}\ee
 We find straightforwardly 
 $$ u_L=8\ka\mu_L  \sin{\rho_L}\frac{ \sin{ \left(\frac{\rho_R}{2}\right)}-\mu_R\cos{\left(\frac{\rho_R}{2}\right)} }{{\sin{\rho_L}\sin{\rho_R}+64\mu_L\mu_R\ka^2\Omega^t\Omega }}\Omega^t(R^2+1)x_R+$$
 \be +\left(\sin{\left(\frac{\rho_L}{2}\right)}+\mu_L\cos{\left(\frac{\rho_L}{2}\right)}\right)\left( \frac{ 1}{1-\mu_LR}R^2 -\frac{ \sin{\rho_L}\sin{\rho_R}  }{\sin{\rho_L}\sin{\rho_R}+64\mu_L\mu_R\ka^2\Omega^t\Omega  }(R^2+1)\right)x_L;\label{568} \ee
 $$ u_R=8\ka\mu_R\sin{\rho_R}\frac{ \sin{ \left(\frac{\rho_L}{2}\right)}+\mu_L\cos{\left(\frac{\rho_L}{2}\right)} }{{\sin{\rho_L}\sin{\rho_R}+64\mu_L\mu_R\ka^2\Omega^t\Omega  }}\Omega(R^2+1)x_L+$$
 \be +\left(-\sin{\left(\frac{\rho_R}{2}\right)}+\mu_R\cos{\left(\frac{\rho_R}{2}\right)}\right)\left( \frac{ 1}{1-\mu_RR}R^2 -\frac{ \sin{\rho_L}\sin{\rho_R}  }{\sin{\rho_L}\sin{\rho_R}+64\mu_L\mu_R\ka^2\Omega^t\Omega  }(R^2+1)\right)x_R.\label{570} \ee

By   inserting $x_L:=\partial_+k_Lk_L^{-1}+k_LA_+k_L^{-1}$ and $x_R:=\partial_+k_Rk_R^{-1}+k_RA_+k_R^{-1}$   into Eqs.\eqref{568},\eqref{570} and in the left-hand-side of Eq.\eqref{565}, then by substituting the result into the right-hand-side of Eq.\eqref{565}, we proceed to the straightforward evaluation of the action \eqref{509}. Indeed, we set $m=(k_L^{-1},k_R^{-1})$, we change the sign of $A$ and
we find that the action \eqref{509} becomes
 $$ S[k_L,k_R,A]=-\int d\tau\oint \sum_{a,b=L,R}{\rm tr}\left((\partial_+ k_a k_a^{-1} -k_aA_{a+}k_a^{-1})M_{ab}(\partial_- k_b k_b^{-1} -k_bA_{b-}k_b^{-1})\right)+$$$$+\ka  \left(I_{\rm WZ}(k_L) - I_{\rm WZ}(k_R)\right)$$\be -\ka\int d\tau\oint {\rm tr}\left(A_-( k_L^{-1}\partial_+ k_L-k_R^{-1}\partial_+ k_R) - A_+( k_L^{-1}\partial_- k_L-k_R^{-1}\partial_- k_R)\right) ,\label{588}\ee
where  $A=A_L=A_R$,  $A_\pm=A_\tau\pm A_\sigma$ are the light cone components of $A=A_\tau d\tau +A_\sigma d\sigma$ and the quantities $M_{ab}$, $a,b=L,R$ are given by
 $$M_{LL}=\frac{2\mu_L\ka}{\sin{\rho_L}(1+\mu_L^2)}\biggl(1+\left(\mu_L^2\cos^2{\frac{\rho_L}{2}}-\sin^2{\frac{\rho_L}{2}}\right)\Bigl(\frac{R}{\mu_L}+\left(1 -64\frac{1+\mu_L^2}{\mu_L\sin{\rho_L}\sin{\rho_R}}\ka^2\mu_R  \Omega^t\tilde \cN^{-1}\Omega\Bigr)(R^2+1)\right)\biggr),$$
 $$M_{RR}=\frac{2\mu_R\ka}{\sin{\rho_R}(1+\mu_R^2)}\biggl(1 +\left(\mu_R^2\cos^2{\frac{\rho_R}{2}}-\sin^2{\frac{\rho_R}{2}}\right)\Bigl(\frac{R}{\mu_R}+\left(1 -64\frac{1+\mu_R^2}{\mu_R\sin{\rho_L}\sin{\rho_R}}\ka^2\mu_L  \Omega \cN^{-1}\Omega^t\Bigr)(R^2+1)\right)\biggr),$$
 $$ M_{LR}=- \left(\sin{\left(\frac{\rho_L}{2}\right)}+
 \mu_L\cos{\left(\frac{\rho_L}{2}\right)}\right)
 \left(\sin{\left(\frac{\rho_R}{2}\right)}+
 \mu_R\cos{\left(\frac{\rho_R}{2}\right)}\right)\frac{16\ka^2}{\sin{\rho_L}\sin{\rho_R}} \cN^{-1}\Omega^t(R^2+1); $$\be
 M_{RL}= \left(\sin{\left(\frac{\rho_R}{2}\right)}-
 \mu_R\cos{\left(\frac{\rho_R}{2}\right)}\right)
 \left(\sin{\left(\frac{\rho_L}{2}\right)}-
 \mu_L\cos{\left(\frac{\rho_L}{2}\right)}\right)\frac{16\ka^2}{\sin{\rho_L}\sin{\rho_R}}\tilde\cN^{-1}\Omega(R^2+1)\label{599}\ee
with
  \be \cN=1+ \frac{64\mu_L\mu_R\ka^2}{\sin{\rho_L}\sin{\rho_R}}\Omega^t\Omega,\quad \tilde\cN=1+\frac{64\mu_L\mu_R\ka^2}{\sin{\rho_L}\sin{\rho_R}}\Omega\Omega^t.\label{558}\ee
It is straightforward to verify
that  the quantities $M_{ab}$, $a,b=L,R$ given by Eqs.\eqref{599} coincide with those  defined in Eqs. \eqref{484}, \eqref{621} and \eqref{453} if we make the following identifications
  \be   \frac{2\kappa}{\sin{\rho_L}}=\frac{1+\mu_L^2}{\mu_L},\quad \frac{2\kappa}{\sin{\rho_R}}=\frac{1+\mu_R^2}{\mu_R},\label{668}\ee\be \Omega=-\frac{1}{4} \cos{\left(\frac{\rho_L}{2}\right)}\cos{\left(\frac{\rho_R}{2}\right)}\omega,\quad \eta_a^2=\mu_a^2\cos^2{\frac{\rho_a}{2}}-\sin^2{\frac{\rho_a}{2}}, 
\quad \A_a=\mu_a\cos^2{\frac{\rho_a}{2}}-\mu_a^{-1}\sin^2{\frac{\rho_a}{2}}, \quad a=L,R.\label{605}\ee
  Assuming those identifications and integrating away $A_\pm$ from \eqref{588}, we recover  the original DHKM model \eqref{438}
 with $k=k_Lk_R^{-1}$. If we do not constrain the parameters by the identifications \eqref{668} and \eqref{605}, then the integrating away of $A_\pm$   gives the version of the  DHKM model with the emancipated parameter $\alpha$. In particular, for the vanishing TsT matrix, we obtain the bi-YB-WZ model \eqref{210c} with the emancipated parameter $\alpha$. The reader may ask now: but where we see in the $\E$-model construction the parameter $\alpha$?  Or, said more precisely, how $\alpha$ is related to the $\E$-model parameters $\mu_L$, $\mu_R$, $\rho_L$, $\rho_R$ and $\ka$?  
 
 To answer these questions, we 
 have to perform a detailed account  of  the number and of the range of the  independent  parameters featuring in  the dressing coset action \eqref{438}. If the identifications \eqref{668} and \eqref{605} are imposed, then the resulting dressing coset \eqref{438}  is characterized by $3+r^2$ parameters $\ka$, $\eta_L$, $\eta_R$ and the TsT matrix $\omega$. Recall that $\ka$ must be an integer in order  that the WZW term ambiguity be the multiple of $2\pi$ and must be positive  in order that the Hamiltonian of the model be positive. To insure also the integrability, the authors of \cite{DHKM} have shown that it must further hold
\be 1+\eta_a^2\geq \ka^2,\qquad a=L,R.\label{677b}\ee

If the identifications \eqref{668} and \eqref{605} are not imposed, we have seemingly
$5+r^2$ free parameters: the same positive integer $\ka$ as before, the TsT matrix $\Omega$ and the real parameters $\mu_L$, $\mu_R$, $\rho_L$, $\rho_R$.  Why do we say "seemingly"? Because   one of those $5+r^2$ parameters turns out to be  superfluous\footnote{The suspicion  that the superfluity of one of the parameters takes place was brought into our mind by the numerical analysis of  Gleb Kotousov \cite{Kot}, who kindly accepted our request to run  the   $SU(2)$ case on the computer.} and there are in reality  just
$4+r^2$ free parameters: $\ka$, $\Omega$, $\rho_L,\rho_R$,  and the sought parameter $\alpha\in]-1,1[$
given by
  \be \al=\frac{ \mu_R-\tan{\left(\frac{\rho_R}{2}\right)}}{ \mu_R +\tan{\left(\frac{\rho_R}{2}\right)}}.\frac{ \mu_L -\tan{\left(\frac{\rho_L}{2}\right)}}{ \mu_L+\tan{\left(\frac{\rho_L}{2}\right)}}.\label{685}\ee
Said in other words, it turns out that in  the resulting $\sigma$-model action \eqref{438}  extracted from the $\E$-model data $D,\E,F$ the parameters $\mu_L$ and
  $\mu_R$ appear only in the combination
  \eqref{685} (this combination can assume any value from the interval $]-1,1[$).

  We provide  two ways  of proving the fact that  if we change $\mu_L$ and $\mu_R$   in  such a way that $\alpha$ does not change then we obtain from those changed $\E$-model data  the same dressing coset $\sigma$-model as from the unchanged ones. 
The first way 
  uses the result
  obtained in Ref.\cite{KS96b}, where it is stated that the Lagrangian of the dressing cosets $\sigma$-models depends only on the parameters characterizing the subspace $V_+^\perp\oplus\F$ of the double $\D$ (cf. Section 3.3 for the notation). In our particular case \eqref{498a},
  \eqref{480a}, we find
  \be V_+^\perp\oplus\F=\left\{u- \frac{\al\cos{(\rho_R)}-\cos{(\rho_L)}}{2\sin{(\rho_R)}\sin{(\rho_L)}}v-
  \ri\frac{1}{\sin{(\rho_R)}}v \ \oplus \  u+ \frac{\al\cos{(\rho_R)}-\cos{(\rho_L)}}{2\sin{(\rho_R)}\sin{(\rho_L)}}v+\ri\frac{\al}{\sin{(\rho_L)}}v;\ u,v\in\K \right\}. \label{697a}\ee
  The second method is more straightforward
and it amounts to the substitution of the formulas \eqref{599} into Eqs.\eqref{446} and then into \eqref{438}.  This   tedious  calculation permits to extract  the   formula \eqref{685} from the explicit form of the obtained $\sigma$-model Lagrangians.  In particular, for the case of the vanishing TsT matrix $\Omega$, the second order action \eqref{438} of the dressing coset  \eqref{588}, \eqref{599}
  becomes simply the action \eqref{210c}
 \be S_{\rm bi-YB-WZ}(k)= \kappa I_{\rm WZ}(k)  + \kappa \int d\tau \oint \  \tr k^{-1}\partial_+k \frac{\al+e^{\rho_LR_k}e^{\rho_RR}}{\al-e^{\rho_LR_k}e^{\rho_RR}}  k^{-1}\partial_-k.\label{653a}\ee
 Recall that $R$ is the Yang-Baxter operator and 
 $R_k$ stands for the operator Ad$_{k^{-1}}R$Ad$_k$.  
 As we already said in Section 1, the action \eqref{653a} may be interpreted as the result of the bi-Yang-Baxter deformation of the WZW model, because the case $\al=0$ corresponds exactly to the WZW model with the level $\ka$.

 Let us say more about the ranges of the original DHKM parameters $\ka$, $\eta_L$, $\eta_R$ and of the dressing coset ones $\ka$, $\rho_L$, $\rho_R$ and $\alpha$.  
 We remark, in particular, that the middle formula of the identification \eqref{605} implies the inequalities
 \be \vert\mu_a\vert\geq \left\vert\tan{\frac{\rho_a}{2}}\right\vert,\qquad q=L,R, \label{707}\ee
 because otherwise the quantities $\eta^2_L$ and $\eta^2_R$ would be negative. We observe also, that if the inequalities \eqref{707} are satisfied then the required inequalities \eqref{677b} hold true as they should.
 Thus we conclude that in the regime $ \vert\mu_a\vert\geq \left\vert\tan{\frac{\rho_a}{2}}\right\vert$ we can reach the original DHKM parametrization by imposing the identifications \eqref{668} and \eqref{605}. In this case, we find easily 
 that it holds
  \be   \alpha=\frac{\eta_L^2+1-\ka}{\eta_L^2+1+\ka}\ \!\frac{\eta_R^2+1-\ka}{\eta_R^2+1+\ka},\qquad  \tan{\frac{\rho_a}{2}}=\frac{\ka\eta_a^2}{\A_a(1+\eta_a^2)},\quad a=L,R.\label{727a}\ee
  Moreover, combining Eqs.\eqref{668} and \eqref{685}, we observe
 that after the identifications \eqref{668} and \eqref{605} are imposed the parameter $\alpha$ is no longer free because it can be expressed just  in terms of $\ka$, $\rho_L$ and $\rho_R$.  
 By the way, the inequalities $ \vert\mu_a\vert\geq \left\vert\tan{\frac{\rho_a}{2}}\right\vert$ imply also that $\alpha$  is  non-negative.  Is there something wrong 
 with the values of $\vert\mu_a\vert$ smaller than
 $\left\vert\tan{\frac{\rho_a}{2}}\right\vert$? No, there is not. From the point of view of our degenerate $\hat\E$-model construction those values are perfectly legitimate and, as we are going to show in the next section, they are also compatible with the
 integrability. Actually, 
 the strictly negative values of $\al$ are reached when $\vert \mu_L\vert < \left\vert\tan{\frac{\rho_L}{2}}\right\vert$  and, simultaneously, $ \vert\mu_R\vert > \left\vert\tan{\frac{\rho_R}{2}}\right\vert$  or vice versa.  If either  $ \mu_L=\tan{\frac{\rho_L}{2}}$ or $ \mu_R=\tan{\frac{\rho_R}{2}}$, then $\alpha=0$.

 Coming back to the case where we do not impose the DHKM identifications \eqref{668} and \eqref{605}, we observe that $\alpha$ is a free parameter independent on $\ka$, $\rho_L$ and $\rho_R$ and that it  takes values in the interval $]-1,1[$. Note however that once  the identifications \eqref{668} and \eqref{605} are applied, $\alpha$ cannot be negative.  What happens
 however in the region $\alpha\leq 0$? Is it so different
 than the region $\alpha\geq 0$? We conjecture that the
 dynamics of the  dressing coset \eqref{438} where the DHKM identifications \eqref{668} and \eqref{605} are not applied  does not change much when we flip the sign of $\alpha$ because the regions $\alpha\geq 0$ and $\alpha\leq 0$ are probably related by some new Poisson-Lie T-duality. The point is that our Drinfeld double $D$ may have more maximally isotropic subgroups than those described by Eq.\eqref{493}, and this fact would lead to a richer T-duality pattern than just that which amounts to the changing of the TsT matrix.
 We
 would make the present paper too voluminous if we wanted to give here the full account of the all Poisson-Lie T-dualities of the DHKM model  (although it is the task which should be accomplished in the future), nevertheless we give here an indication why we believe  that the regions $\alpha\geq 0$ and $\alpha\leq 0$ are  related by some new Poisson-Lie T-duality. It is because it is true in the limiting case $\rho_L\to 0$.  
 
Taking the limit $\rho_L\to 0$ at the $\E$-model level is a subtle exercise because of the singularity of the bilinear form $(.,.)_\D$ given by Eq.\eqref{498a}, however, this limit can be easily considered for the second order action \eqref{653a}; we obtain simply
 \be S_{\rm YB-WZ}(k)= \kappa I_{\rm WZ}(k)  + \kappa \int d\tau \oint \  \tr k^{-1}\partial_+k \frac{\al+ e^{\rho_RR}}{\al- e^{\rho_RR}}  k^{-1}\partial_-k.\label{731}\ee
As already the notation indicates,   the $\sigma$-model \eqref{731} turns out to be   nothing but the so called YB-WZ model introduced in Ref.\cite{DMV15}. To see it, we rewrite \eqref{731} as
 \be S_{\rm YB-WZ}(k)= \kappa I_{\rm WZ}(k)  + \kappa \int d\tau \oint  \  \tr k^{-1}\partial_+k (\mathbf a +\mathbf b R+\mathbf c R^2)k^{-1}\partial_-k,\label{733}\ee
 where 
 \be \mathbf a=\frac{\al+1}{\al-1}, \quad \mathbf b=\frac{2\alpha\sin{\rho_R}}{1-2\al\cos{\rho_R}+\al^2}, \quad \mathbf c=\frac{\al+1}{\al-1}+\frac{1-\al^2}{1-2\al\cos{\rho_R}+\al^2}.\label{739}\ee
 It can be checked easily, that it holds
 \be \mathbf b^2=\frac{\mathbf c}{\mathbf a}(\mathbf a^2-\mathbf a\mathbf c-1),\label{737}\ee
 therefore our $\sigma$-model \eqref{731} coincides with 
 the YB-WZ model as described in Ref.\cite{DDST}. Moreover, it was shown it \cite{DDST}, that the model \eqref{733} supplemented by the constraint  \eqref{737} is Poisson-Lie T-dual to the model 
  \be \tilde S_{\rm YB-WZ}(k)= \kappa I_{\rm WZ}(k)  + \kappa \int d\tau \oint  \  \tr k^{-1}\partial_+k (\tilde {\mathbf a}+\tilde {\mathbf b}R+\tilde {\mathbf c}R^2)k^{-1}\partial_-k,\label{740}\ee
  where \be \tilde {\mathbf a}=\frac{1}{{\mathbf a}}, \quad \tilde{\mathbf b}=-{\mathbf b},\quad \tilde {\mathbf c}=-\frac{{\mathbf b}^2}{{\mathbf c}}  \label{741}\ee
 and
  \be \qquad \tilde {\mathbf b}^2=\frac{\tilde {\mathbf c}}{\tilde {\mathbf a}}(\tilde {\mathbf a}^2-\tilde {\mathbf a}\tilde {\mathbf c}-1).\label{749}\ee
  Rewriting the T-dual $\sigma$-model \eqref{740} back into our form as 
   \be \tilde S_{\rm YB-WZ}(k)= \kappa I_{\rm WZ}(k)  + \kappa \int d\tau \oint  \  \tr k^{-1}\partial_+k \frac{\tilde\al+ e^{\tilde\rho_RR}}{\al- e^{\tilde\rho_RR}}  k^{-1}\partial_-k,\label{745}\ee
   we find that the T-duality transformation \eqref{741} gets translated into
   \be \tilde \al=-\al,\qquad \frac{1+\tilde \al}{1-\tilde\al}\tan{\frac{\tilde\rho_R}{2}}=
   \frac{1+\al}{1-\al}\tan{\frac{\rho_R}{2}}.\ee
   Said in other words, the T-duality indeed exchanges the regions  $\alpha\geq 0$ and $\alpha\leq 0$.
   
   What happens if the value of $\alpha$ in the bi-YB-WZ action \eqref{653a}  approaches the borders of the interval $]-1,1[$? Consider e.g. the case $\al\to 1$. This is a singular limit but it may give rise to a nontrival structure if we let at the same time $\ka$, $\rho_L$ and $\rho_R$ tend to zero. More precisely, we set  
   \be \ \rho_L=2\ka b_L,\quad \rho_R=2\ka b_R, \quad \al=e^{-2\ka  a},\quad  a>0\label{756}\ee
  and then consider the limit $\ka\to 0$ in the action \eqref{653a}. In this way we obtain the   bi-Yang-Baxter deformation of the principal chiral model \cite{K}:
    \be S_{\rm bi-YB}(k)= - \int d\tau \oint d\sigma \  \tr\left( k^{-1}\partial_+k \frac{1}{a+b_R R+b_LR_k }k^{-1} \partial_-k\right).\label{755}\ee

  \section{Integrability of the DHKM model}
  \setcounter{equation}{0}
  The purpose of the present section is to prove the integrability of   the   DHKM model \eqref{588}, \eqref{599} in the case when the DHKM identifications \eqref{668} and \eqref{605} are not imposed. Said in other words, our concern is to study the integrability if the parameter $\alpha$ is emancipated. For that, we shall work directly in the $\E$-model formalism.
     Our strategy of proof
  will consist in  representing the  field equations of the model  in terms of two $\K$-valued currents $J=J_\tau d\tau +J_\sigma d\sigma$ and $B=B_\tau d\tau +B_\sigma d\sigma$
  as follows
  \be \partial_+B_- -\partial_-B_+ +[B_+,B_-]+\xi^2[J_+,J_-]=0,\label{626}\ee
  \be \partial_+J_-+[B_+,J_-]=0,\quad \partial_-J_++[B_-,J_+]=0.\label{627}\ee
  Here $B_\pm=B_\tau\pm B_\sigma$ and $J_\pm=J_\tau\pm J_\sigma$.
As shown in Section 2.1.3 of Ref.\cite{DLMV}, the system of the equations  \eqref{626} and \eqref{627} admits the following Lax pair with
  the spectral parameter $z$
  \be \L_\pm(z)=-B_\pm-\xi z^{\pm 1}J_\pm.\label{650}\ee
  Indeed, it is straightforward to check that the zero curvature condition 
    \be \partial_+\L_-(z)- \partial_-\L_+(z) -[\L_+(z),\L_-(z)]=0\label{633}\ee
 is equivalent to the system \eqref{626} and \eqref{627}.
 
 Coming back to the general dressing coset story  described in Section 3.3, it is easy to work out the field equations
 of the isotropically gauged $\E$-model \eqref{309}. They read
 \be \partial_\tau ll^{-1}-A_\tau=\E(\partial_\sigma ll^{-1}-A_\sigma);\label{657}\ee
 \be (\partial_\sigma ll^{-1}-A_\sigma)_2=0;\label{658}\ee
  \be \left( \partial_\tau ll^{-1}-A_\tau \right)_2=0,\label{659}\ee
  where we recall that every element $y$ of the Drinfeld double Lie algebra $\D$ can be unambiguously written as 
  \be y=y_0+y_1+y_2,\qquad y_0\in V^\perp=(\F\oplus\E\F)^\perp, \quad y_1\in\F, \quad y_2\in\E\F.\label{661}\ee
  In what follows, we shall need a refinement of the decomposition \eqref{661} which is obtained by decomposing further $y_0$ as
  \be y_0=y_+ + y_-,\label{663}\ee
  where 
  \be\E y_\pm=\pm y_\pm.\label{665}\ee
  We write also
  \be y_\pm\in V_\pm^\perp,\qquad V_\pm^\perp=(1\pm\E)V^\perp.\label{667}\ee
With this notation, the equations of motion \eqref{657},\eqref{658} and \eqref{659} can be rewritten as
\be  (\partial_\pm ll^{-1})_2=0,\quad  (\partial_\pm ll^{-1})_1=A_\pm,\quad (\partial_\pm ll^{-1})_\mp=0.\label{669}\ee  
Indeed, the relations $(\partial_\pm ll^{-1})_2=0$
are the direct consequences of the equations
\eqref{658} and \eqref{659}, because $A_\tau,A_\sigma\in\F$.
Furthermore, the fact that $\partial_\sigma ll^{-1}-A_\sigma\subset \F^\perp=V^\perp\oplus\F$ 
and, simultaneously, $\E(\partial_\sigma ll^{-1}-A_\sigma)\subset \F^\perp$
means  that $\partial_\sigma ll^{-1}-A_\sigma\subset  V^\perp$, or, said differently, that $(\partial_\sigma ll^{-1})_1=A_\sigma$. This is true because    $\E V^\perp=V^\perp$ and $\F\cap\E\F=\{0\}$ so that a
non-vanishing component $(\partial_\sigma ll^{-1}-A_\sigma)_1$ would imply  the non-vanishing component
of $(\E(\partial_\sigma ll^{-1}-A_\sigma))_2$.
Similarly, using the fact that the operator $\E$ squares to the identity we can rewrite Eq.\eqref{657} as
 \be \E(\partial_\tau ll^{-1}-A_\tau)=\partial_\sigma  ll^{-1}-A_\sigma\label{676}\ee
 and then the changing the roles of $\sigma$ and $\tau$ in the previous chain of arguments leads to the conclusion that $(\partial_\tau ll^{-1})_1=A_\tau$. In this way, we have proved that
  $(\partial_\pm ll^{-1})_1=A_\pm$.
  Finally, knowing that $(\partial_\pm ll^{-1})_2=0$
  and $(\partial_\pm ll^{-1})_1=A_\pm$ permits to rewrite Eq.\eqref{657} as
  \be (\partial_\tau ll^{-1})_0 =\E(\partial_\sigma ll^{-1})_0,\label{684}\ee
  or, equivalently, as 
  \be \E(\partial_\tau ll^{-1})_0=(\partial_\sigma ll^{-1})_0.\label{686}\ee
  Adding and substracting Eqs.\eqref{684} and \eqref{686} gives $(\partial_\pm ll^{-1})_\mp=0$.
  
  We now specify the equations of motions \eqref{669} for the gauged $\E$-model constructed in Section 4, which underlies the $(4+r^2)$-parametric   DHKM model \eqref{588}, \eqref{599} with the identifications \eqref{668} and \eqref{605} not imposed. Recall that the Drinfeld double $D$ is in this case the direct product $K^\bc\times K^\bc$, the invariant bilinear form $(.,.)_\D$ on the Lie algebra $\D=\K^\bc\oplus \K^\bc$ is given by the formula
 \be \left(z_L\oplus z_R,z'_L \oplus z'_R\right)_\D:=\frac{4\ka}{\sin{(\rho_L)}}  \Im\tr\left(e^{\ri \rho_L}z_Lz'_L\right)+\frac{4\ka}{\sin{(\rho_R)}}\Im\tr\left(e^{-\ri \rho_R}z_Rz'_R\right),\quad z_a,z'_a\in \K^\bc, \ a=L,R, \label{830}\ee
  the isotropic subalgebra $\F$ consists of the 
 elements $(x\oplus x)$, $x\in\K$
  and  the eponymous self-adjoint operator $\E:\D\to\D$ commuting with the adjoint action of $\F$ is given by the formula
\be \E(z_L\oplus z_R)= \ri\left(\frac{1-\mu_L^2}{2\mu_L}z_L+e^{-\ri \rho_L}\frac{1+\mu_L^2}{2\mu_L}z_L^*\right)\oplus  \ri\left(\frac{1-\mu_R^2}{2\mu_R}z_R+e^{\ri \rho_R}\frac{1+\mu_R^2}{2\mu_R}z_R^*\right). \label{837d}\ee
We want to describe the subspaces $V^\perp_\pm\subset\D$ corresponding to those data.
We note that $V^\perp_\pm\subset ({\rm Id}\pm\E)\D$, where the half-dimensional subspaces $ ({\rm Id}\pm\E)\D$ can be conveniently parametrized in terms of the elements $u_{L,R}^{\pm}$ of the Lie algebra $\K$ as
\be  ({\rm Id}\pm\E)\D=\{e^{\ri\left(\mp m_L-\jp\rho_L\right)}u_L^\pm\oplus e^{\ri\left(\mp m_R+\jp\rho_R\right)}u_R^\pm, \quad u_{L,R}^{\pm}\in \K\}, \label{697}\ee
where we have traded the parameters $\mu_{L,R}$ for 
$m_{L,R}$ according to the formulas
\be \mu_a=\tan{(m_a)},\quad a=L,R.\label{700}\ee
Note that the subspaces
$V_\pm^\perp$ are formed by the vectors in $ ({\rm Id}\pm\E)\D$
  perpendicular to $\F$ which gives
  \be V_\pm^\perp =\{e^{\ri\left(\mp m_L-\jp\rho_L\right)}\frac{\sin{\left(\pm m_R+\frac{\rho_R}{2}\right)}}{\sin{(\rho_R)}}J_\pm\oplus
  e^{\ri\left(\mp m_R+\jp\rho_R\right)}\frac{\sin{\left(\mp m_L+\frac{\rho_L}{2}\right)}}{\sin{(\rho_L)}}J_\pm,\quad J_\pm\in\K\}.\label{743}\ee
The equations of motions \eqref{669} are then solved by
\be \d_+ll^{-1}=e^{-\ri\left( m_L+\jp\rho_L\right)}\frac{\sin{\left( m_R+\frac{\rho_R}{2}\right)}}{\sin{(\rho_R)}}J_+\oplus
  e^{-\ri\left( m_R-\jp\rho_R\right)}\frac{\sin{\left( -m_L+\frac{\rho_L}{2}\right)}}{\sin{(\rho_L)}}J_++A_+\oplus A_+,\label{708}\ee
 \be \d_-ll^{-1}=e^{\ri\left( m_L-\jp\rho_L\right)}\frac{\sin{\left( -m_R+\frac{\rho_R}{2}\right)}}{\sin{(\rho_R)}}J_-\oplus
  e^{\ri\left( m_R+\jp\rho_R\right)}\frac{\sin{\left( m_L+\frac{\rho_L}{2}\right)}}{\sin{(\rho_L)}}J_-+A_-\oplus A_-,\label{710}\ee
  where the $\K$-valued fields $J_\pm$, $A_\pm$ must be such that the Bianchi identity be verified:
  \be \d_-(\d_+ll^{-1})-\d_+(\d_-ll^{-1})+[\d_+ll^{-1}, \d_-ll^{-1}]=0.\label{712}\ee
  Inserting the expressions \eqref{708}, \eqref{710} into the left-hand-side of \eqref{712}, we obtain 
  \be \Bigl(\d_-(\d_+ll^{-1})-\d_+(\d_-ll^{-1})+[\d_+ll^{-1}, \d_-ll^{-1}]\Bigr)_2=0;\label{714}\ee
    $$\Bigl(\d_-(\d_+ll^{-1})-\d_+(\d_-ll^{-1})+[\d_+ll^{-1}, \d_-ll^{-1}]\Bigr)_1=$$\be =\Bigl(\d_-A_+-\d_+A_-+[A_+,A_-]+\beta[J_+,J_-]\Bigr)\oplus \Bigl(\d_-A_+-\d_+A_-+[A_+,A_-]+\beta[J_+,J_-]\Bigr)
    ;\label{716}\ee
   $$\Bigl(\d_-(\d_+ll^{-1})-\d_+(\d_-ll^{-1})+[\d_+ll^{-1}, \d_-ll^{-1}]\Bigr)_+=$$
   \be =e^{-\ri\left( m_L+\jp\rho_L\right)}S_+^L\Bigl(\d_-J_+ -[A_-+\alpha_+J_-,J_+]\Bigr)\oplus
  e^{-\ri\left( m_R-\jp\rho_R\right)}S_+^R\Bigl(\d_-J_+ -[A_-+\alpha_+J_-,J_+]\Bigr);\label{719}\ee
  $$\Bigl(\d_-(\d_+ll^{-1})-\d_+(\d_-ll^{-1})+[\d_+ll^{-1}, \d_-ll^{-1}]\Bigr)_-=$$
   \be =e^{\ri\left( m_L-\jp\rho_L\right)}S_-^L\Bigl(-\d_+J_- +[A_++\alpha_-J_+,J_-]\Bigr)\oplus
  e^{\ri\left( m_R+\jp\rho_R\right)}S_-^R\Bigl(-\d_+J_- +[A_++\alpha_-J_+,J_-]\Bigr);\label{722}\ee
  where

  \be \alpha_\pm=\pm\frac{1}{\al-\al^{-1}}\left(\frac{S_\pm^R}{S_\pm^L}-\frac{S_\pm^L}{S_\pm^R}\right),\quad \beta=S_+^LS_-^L-\frac{\cos{(m_L)}}{\cos{\left(\frac{\rho_L}{2}\right)}}(S_+^L\alpha_++S_-^L\alpha_-)\label{766}\ee
    \be S_\pm^L:=\frac{\sin{\left( \pm m_R+\frac{\rho_R}{2}\right)}}{\sin{(\rho_R)}},  \quad S_\pm^R:=\frac{\sin{\left( \mp m_L+\frac{\rho_L}{2}\right)}}{\sin{(\rho_L)}}, \quad \al=\frac{S_-^LS_+^R}{S_+^LS_-^R}\label{767}\ee
    Note that the parameter $\al$ is the same as the one featuring in Eq.\eqref{685}.
 
  The equation \eqref{712} together with Eqs.\eqref{714}, \eqref{716}, \eqref{719} and \eqref{722} then imply the validity of the following system of equations
  \be \d_-J_+ -[A_-+\alpha_+J_-,J_+]=0,\quad \d_+J_- -[A_++\alpha_-J_+,J_-]=0;\label{728}\ee
  \be \d_-A_+-\d_+A_-+[A_+,A_-]+\beta[J_+,J_-]=0.\label{729}\ee
  The system \eqref{728} and \eqref{729} admits the Lax pair \eqref{650} with  spectral parameter because it is equivalent to the system \eqref{626} and \eqref{627} upon the identification
  \be B_\pm=-A_\pm -\alpha_\mp J_\pm, \quad \xi^2=\alpha_+\alpha_-+\beta.\ee
  The    dressing coset  \eqref{588}, \eqref{599} is therefore integrable also in the case when the DHKM identifications \eqref{668} and \eqref{605} are not imposed.
  
  \section{Renormalization of the   DHKM model}
    \setcounter{equation}{0}
\subsection{Generalities about the renormalization of the non-degenerate $\E$-models}

 The fact that a given $\sigma$-model has the first order dynamics which can be expressed in terms of a non-degenerate $\E$-model is very
 useful for the study of its ultraviolet properties, because such model is automatically renormalizable. Indeed, it was established in 
 \cite{VKS,SfS,SSD10}, that the renormalization group flow respects the structure of the $\E$-model, just flowing from one epynomous operator
 $\E$ to another. This flow is described by an elegant formula derived in \cite{SSD10} (and used in an different context already in \cite{T,FR}):
 \be \frac{d\E_{AB}}{ds}=\left(\E_{AC}\E_{BF}-\eta_{AC}\eta_{BF}\right)\left(\E^{KD}\E^{HE}-\eta^{KD}\eta^{HE}\right)f_{KH}^{\ \ \ \ \!C}f_{DE}^{\ \ \ \ \!F}.\label{Sf}\ee
Here $s$ is the RG flow parameter and the capital Latin indices refer to the choice of a basis $T_A$ of the Lie algebra $\D$:
\be \E_{AB}:=(T_A,\E T_B)_\D, \quad \eta_{AB}:=(T_A,T_B)_\D,\quad [T_A,T_B]=f_{AB}^{\phantom{AB}C}T_C.\ee
The indices are lowered  and raised with the help of the tensor $\eta_{AB}$ and its inverse.

Up to an irrelevant normalization constant, the flow formula \eqref{Sf} can be cast in the basis-independent way   \cite{K18}:
\be \frac{d\E}{ds}=\P_+[[\P_+,\P_-]]\P_-+\P_-[[\P_+,\P_-]]\P_+,\label{flo}\ee
where the projections $\P_\pm$ are defined as
\be \P_\pm=\jp(1\pm \E),\ee
and the bracket $[[.,.]]:S^2\D\times S^2\D\to S^2\D$  is defined on the symmetric product $S^2\D$ as
\be [[A,B]]:=[A',B']\otimes [A'',B''].\ee
Here we use the Sweedler notation $A=A'\otimes A''$, $B=B'\otimes B''$
and we view the self-adjoint operators $\P_\pm$ as the elements of $S^2\D$
in the sense of the formula
\be \P_\pm x:=\P_\pm'(\P''_\pm,x)_\D, \quad x\in \D.\ee

We give now an equivalent description of the
RG flow of the operator $\E$ in terms of the flow
of the half-dimensional subspaces $E_\pm\subset \D$ constituted by the eigenvectors of $\E$ corresponding to the eigenvalue $\pm 1$. The formula \eqref{flo} is then equivalent to the following infinitesimal changes of the vector spaces $E_\pm$:
\be E_\pm +\delta E_\pm=\{v_\pm +\frac{ \delta s}{2}\S_\pm v_\pm,\ v_\pm\in E_\pm\},\label{806a}\ee
where the operators $\S_\pm:E_\pm\to E_\mp$
are given by the formulas
\be \S_\pm:=\pm  \P_\mp[[\P_+,\P_-]]\P_\pm.\ee
The advantage to work with the operators 
$\S_\pm$ is technical, because the following  simple formulas can be straightforwardly  derived for their matrix elements \cite{SV} :  
\be (v_\mp, \S_\pm v_\pm)_\D=\mp\tr(\P_\pm\ad_{v_\pm}\P_\mp\ad_{v_\mp}\P_\pm), \quad v_\pm\in E_\pm.\label{812}\ee
Let us illustrate how the   formula  \eqref{812} accounts for the renormalization of the Yang-Baxter $\sigma$-model
\cite{K}, the action of which reads
 \be S_{\rm YB}(k)=  -\int d\tau \oint d\sigma \  \tr\left( k^{-1}\partial_+k \frac{1}{a+b R }k^{-1} \partial_-k\right).\label{815}\ee
 Here $a>0$ and $b\geq 0$ are real  parameters, tr is negative definite and $R$ is the Yang-Baxter operator.
 
 The $\E$-model underlying the Yang-Baxter $\sigma$-model was first constructed in
 \cite{K}. The Drinfeld double $D$ is the complexification $K^\bc$ of the simple compact Lie group $K$, the invariant bilinear form $(.,.)_\D$ on the Lie algebra $\D$ is given by the formula
 \be (z,w)_\D=\frac{2}{b}\Im\tr(zw),\quad z,w\in\K^\bc\ee
 and the operator $\E$ reads
 \be \E z=\frac{\ri}{2}\left(\left( \frac{a}{b}-\frac{b}{a}
\right)z+
\left(\frac{b}{a}+\frac{a}{b}\right)z^*\right).\ee
If $x_\pm\in\K$ then 
\be \P_\pm x_\pm= \jp\left(1\mp\ri\frac{b}{a}\right)x_\pm.\ee
Picking $u\in\K$, we find easily  
\be \P_+\ad_{\P_+x_+}\P_-\ad_{\P_-x_-}\P_+u
=\js\left(1+\frac{b^2}{a^2}\right)\P_+\ad_{\P_+x_+}\P_-[x_-,u]=\frac{1}{16}\left(1+\frac{b^2}{a^2}\right)^2\P_+[x_+,[x_-,u]],\ee
hence
\be \tr\left(\P_+\ad_{\P_+x_+}\P_-\ad_{\P_-x_-}\P_+\right)=\frac{1}{16}\left(1+\frac{b^2}{a^2}\right)^2\tr\left(\ad_{x_+}\ad_{x_-}\right)\equiv \frac{c_K}{16}\left(1+\frac{b^2}{a^2}\right)^2\tr(x_+x_-).\ee
Note that $c_K$ is the double of the dual Coxeter number (for example, $C_K=4$ for $su(2)$).

At the same time, we have for the RG flow
\[\begin{aligned}\delta\left(\frac{b}{a}\right)&=\frac{ 
b(\P_-x_-, \ri x_+)_\D}{\tr(x_+x_-)}\delta\left(\frac{b}{a}\right)= -\frac{ 
2b(\P_-x_-, \delta(\P_+x_+))_\D}{\tr(x_+x_-)}=-\frac{ 
b(\P_-x_-, \S_+(\P_+x_+))_\D}{\tr(x_+x_-)}\delta s=\\&=\frac{ b\ \!\tr\left(\P_+\ad_{\P_+x_+}\P_-\ad_{\P_-x_-}\P_+\right)}{\tr(x_+x_-)}\delta s=\frac{bc_K}{16}\left(1+\frac{b^2}{a^2}\right)^2\delta s.\end{aligned}\]
The match of this formula with the RG flow of the Yang-Baxter $\sigma$-model   obtained  in the literature is perfect (cf. Eq. (4.9) of Ref. \cite{SST15}  with the identification of the parametres: $\zeta=0$, $t=a$, $t\eta=b$).
Note also, that $b$ does not flow, being the kinematical parameter characterizing the inner product $(.,.)_\D$; the flow of the
parameter $a$ is therefore
\be \dot a=-\frac{c_K}{16a^2}(a^2+b^2)^2 .\ee

 \subsection{Generalities about the renormalization of the degenerate $\E$-models}
 
The automatic renormalizability  of the degenerate $\hat\E$-models (the dressing cosets) was established  in \cite{SV}. As explained in Section 3.2, the degenerate $\hat\E$-model is characterized by the $\br$-linear self-adjoint operator $\hat\E:\F^\perp\to\F^\perp$, that commutes with the adjoint action of the isotropic subalgebra $\F$  on $\F^\perp$ and verifies few other properties, that is, its kernel must contain $\F$, the image of the operator $\hat\E^2-{\rm Id}$ has to be contained in $\F$ and the bilinear form $(.,\hat\E .)_{\F^\perp}$ has to be positive semi-definite. The description of the RG flow of the operator $\hat\E$ is done in \cite{SV} in terms of the flow of the 
subspaces $\hat E_\pm\subset \D$ defined as
\be \hat E_\pm={\rm Im}(\hat\E^2\pm\hat\E).\ee
It is fully sufficient to consider just the flow of the subspace $\hat E_+$   which 
is
\be \hat E_+ +\delta \hat E_+=\{\hat v_+ +\frac{\delta s}{2} \hat \S_+\hat v_+,\ \hat v_+\in \hat E_+\},\label{848}\ee
where the operator $\hat\S_+:\hat E_+\to \hat E_-$
is  characterized by its matrix elements
\be (\hat v_-, \hat\S_+ v_+)_\D=-\tr(\hat\P_+\ad_{\hat v_+}\hat\P_+^\perp\ad_{\hat v_-}\hat\P_+), \quad \hat v_\pm\in \hat E_\pm.\label{851}\ee
We denote  $\hat P_\pm$  the orthogonal projections on the subspaces $\hat E_\pm$ and 
$\hat \P_+^\perp={\rm Id}_\D-\hat\P_+$. Note
that $\hat \P_+^\perp$ {\it does not}  project on the subspace $\hat E_-$! Only in the case  $\F=\{0\}$, that is in the non-degenerate case, the projections $\hat \P_+^\perp$ and
$\hat \P_-$ get equal, the non-degenerate formula \eqref{812} is then the special case of the general formula \eqref{851}.

Let us now illustrate how the  formula  \eqref{812} accounts for the renormalization of the bi-Yang-Baxter deformation of the principal chiral model introduced in 
\cite{K}.  The action of this $\sigma$-model reads
 \be S_{\rm bi-YB}(k)= - \int d\tau \oint d\sigma \  \tr\left( k^{-1}\partial_+k \frac{1}{a+b_R R+b_LR_k }k^{-1} \partial_-k\right),\label{861}\ee
 where $a>0$, $b_L\geq 0$ and $b_R\geq 0$  are real  parameters, tr is negative definite,  $R$ is the Yang-Baxter operator and  $R_k$ stands for the operator Ad$_{k^{-1}}R$Ad$_k$.
 The RG flow of the parameters $a,b_L,b_R$ has been found previously in Ref.\cite{SST15} and it reads
 \be \dot b_R=\dot b_L=0,\quad \dot a=-\frac{c_K}{16a^2}(a^2+(b_L-b_R)^2)(a^2+(b_L+b_R)^2).\label{865}\ee
 Thus our aim is to recover the flow \eqref{865} from the formula \eqref{851}.
 
Let us first interpret
 the $\sigma$-model action \eqref{861} as   the  dressing coset. The relevant Drinfeld double $D$ turns out to be  the direct product $K^\bc\times K^\bc$, the invariant bilinear form $(.,.)_\D$ on the Lie algebra $\D=\K^\bc\oplus \K^\bc$ is given by the formula
 \be \left(z_L\oplus z_R,z'_L \oplus z'_R\right)_\D:=\frac{2}{b_L}\Im\tr\left( z_Lz'_L\right)+\frac{2}{b_R}\Im\tr\left( z_Rz'_R\right),\quad z_a,z'_a\in \K^\bc, \quad a=L,R\label{870}\ee
and  the operator $\E:\D\to\D$  reads
\be   \E(z_L\oplus z_R)= \ri\left(\frac{1-\mu_L^2}{2\mu_L}z_L+ \frac{1+\mu_L^2}{2\mu_L}z_L^*\right)\oplus  \ri\left(\frac{1-\mu_R^2}{2\mu_R}z_R+ \frac{1+\mu_R^2}{2\mu_R}z_R^*\right).  \label{872}\ee
Moreover, the sign of $\mu_L$ must be the same as the sign of $b_L$ and the same thing must be true for $\mu_R$ and $b_R$.

Choosing $\F=\K\oplus \K$, we follow the procedure described in Section 3.3 and obtain the $\sigma$-model \eqref{861} out from the gauged $\E$-model action \eqref{309}. The identification of the parameters is as follows
\be a=\frac{b_L}{\mu_L}+\frac{b_R}{\mu_R}.\ee
Following the general construction of Section 3.3, we identify the subspaces $\hat E_\pm$ with the subspaces $({\rm Id}+\E)V^\perp$, that is
\be \hat E_\pm=\left\{\frac{\mp b_L}{2\mu_L}(1\mp\ri\mu_L)v_\pm\oplus \frac{\pm b_R}{2\mu_R}(1\mp\ri\mu_R)v_\pm;\  v_\pm \in\K\right\}.\ee
Accordingly, it is convenient to parametrize the elements of $\hat E_\pm$ in terms of the Lie algebra $\K$ by constructing the following bijective maps $q_\pm:\K\to\hat E_\pm$:
\be q_\pm(x):=\frac{\mp b_L}{2\mu_L}(1\mp\ri\mu_L)x\oplus \frac{\pm b_R}{2\mu_R}(1\mp\ri\mu_R)x.\ee
Now we pick $x_\pm,u\in\K$ and we find   
\[\begin{aligned} &\ \hat\P_+\ad_{q_+(x_+)}\hat\P_+^\perp\ad_{q_-(x_-)}\hat \P_+q_+(u)
=-\js\hat\P_+\ad_{q_+(x_+)}\hat\P_+^\perp\left(\frac{b_L^2(1+\mu_L^2)}{\mu_L^2}[x_-,u]\oplus
\frac{b_R^2(1+\mu_R^2)}{\mu_R^2}[x_-,u]\right)\\
&=-\js\hat\P_+\ad_{q_+(x_+)}\left(\frac{b_Lb_R}{a\mu_L\mu_R}\left(a+\mu_Lb_L+\mu_Rb_R\right)([x_-,u]\oplus[x_-,u])+\left(\frac{b_L^2-b_R^2}{a}+\frac{b_L}{\mu_L}
-\frac{b_R}{\mu_R}\right) q_-([x_-,u])\right)=\\&=-  \frac{b_Lb_R}{4a\mu_L\mu_R}\left(a+\mu_Lb_L+\mu_Rb_R\right)q_+([x_+,[x_-,u])-\js\left(\frac{b_L^2-b_R^2}{a}+\frac{b_L}{\mu_L}
-\frac{b_R}{\mu_R}\right) \hat\P_+[q_+(x_+),q_-([x_-,u])]=\\&=-\js\left(\frac{b_Lb_R}{a\mu_L\mu_R}\left(a+\mu_Lb_L+\mu_Rb_R\right)+\js\left(\frac{b_L^2-b_R^2}{a}+\frac{b_L}{\mu_L}
-\frac{b_R}{\mu_R}\right)^2\right) q_+([x_+,[x_-,u]])=\\&=-\frac{1}{16a^2}(a^2+(b_L-b_R)^2)(a^2+(b_L+b_R)^2)\ \!q_+([x_+,[x_-,u]]).
 \end{aligned}\label{885}\]  
 We infer that
 \be \tr\left(\hat\P_+\ad_{q_+(x_+)}\hat\P_+^\perp\ad_{q_-(x_-)}\hat \P_+\right)=-\frac{c_K}{16a^2}(a^2+(b_L-b_R)^2)(a^2+(b_L+
 b_R)^2)\tr(x_+x_-).\ee
At the same time, we have for the RG-flow
\[\begin{aligned} \delta a  &=\frac{ \left(q_-(x_-),\delta\left(\frac{b_L}{\mu_L}\right)x_+\oplus -\delta\left(\frac{b_R}{\mu_R}\right)x_+\right)_\D}{\tr(x_-x_+)}=-\frac{ 
 \Bigl(q_-(x_-),2 \delta q_+(x_+)\Bigr)_\D}{\tr(x_+x_-)}=-\frac{ 
 \left(q_-(x_-), \hat\S_+q_+(x_+)\right)_\D}{\tr(x_+x_-)}\delta s=\\&=\frac{\tr(\hat\P_+\ad_{q_+(x_+) }\hat\P_+^\perp\ad_{q_-(x_-) }\hat\P_+)}{\tr(x_+x_-)}\delta s=-\frac{c_K}{16a^2}(a^2+(b_L-b_R)^2)(a^2+(b_L+b_R)^2)\delta s.\end{aligned}\]
This formula matches perfectly  the RG flow \eqref{865} of the bi-Yang-Baxter deformation of the principal chiral model as  obtained  in the literature (cf. Eq. (4.9) of Ref. \cite{SST15}  with the identification of the parametres: $t=a$, $t\zeta=b_L$ and $t\eta=b_R$).

\subsection{Renormalizability of the bi-YB-WZ model}

Now we are coming up to the true concern of the present section which is to establish the renormalization group flow of the bi-YB-WZ model. Thanks to the formula \eqref{851}, we can do that working directly in the first order $\E$-model formalism.  The great advantage of this first order approach resides in the fact that it tells us  immediately which parameters do not flow. Indeed, all parameters that characterize the structure of the Drinfeld double are RG invariant; in the present context, this statement concerns the parameters $\ka$, $\rho_L$ and $\rho_R$ which enter in the definition \eqref{498a} of the bilinear form $(.,.)_\D$. Moreover,  the parameters featuring in the $\sigma$-model action which characterize the embedding of the maximally isotropic subgroup in the double  are not present in the first order $\E$-model data and they therefore neither flow nor they influence the flow of the $\E$-model parameters. In particular,  the TsT matrix appearing 
in \eqref{493} can be safely set to zero without any lack of generality. 

The only parameters which can flow are thus  those which  characterize the 
operator $\E$, or, speaking more precisely, those characterizing the subspace $V_+^\perp\oplus\F$  in the notation of Section 3.3. In the DHKM context, a quick glance  at the formula \eqref{697a} makes us to conclude that the sole parameter which can flow is $\alpha$. This is  a nontrivial statement! Indeed, considering  the bi-YB-WZ action 
 \be S_{\rm bi-YB-WZ}(k)= \kappa I_{\rm WZ}(k)  + \kappa \int d\tau \oint \  \tr k^{-1}\partial_+k \frac{\al+e^{\rho_LR_k}e^{\rho_RR}}{\al-e^{\rho_LR_k}e^{\rho_RR}}  k^{-1}\partial_-k,\label{999}\ee
would we see easily   without the $\E$-model insight 
 that the action  \eqref{999}  is  written in a RG friendly way, that is, only the parameter $\alpha$ can flow   and all other parameters $\ka$, $\rho_L$ and $\rho_R$ are RG invariant?  Of course, this qualitative insight is not enough for us and we are now going to determine
the flow of $\alpha$ quantitatively.

  We start by recalling the DHKM  $\E$-model set up introduced in Section 4. The Drinfeld double $D$ is the direct product $K^\bc\times K^\bc$, the invariant bilinear form $(.,.)_\D$ on the Lie algebra $\D=\K^\bc\oplus \K^\bc$ is given by the formula
\be \left(z_L\oplus z_R,z'_L \oplus z'_R\right)_\D:=\frac{4\ka}{\sin{(\rho_L)}}  \Im\tr\left(e^{\ri \rho_L}z_Lz'_L\right)+\frac{4\ka}{\sin{(\rho_R)}}\Im\tr\left(e^{-\ri \rho_R}z_Rz'_R\right),\quad z_a,z'_a\in \K^\bc, \ a=L,R\label{1007}\ee
and the non-degenerate operator $\E:\D\to\D$  is given by  
\be \E(z_L\oplus z_R)= \ri\left(\frac{1-\mu_L^2}{2\mu_L}z_L+e^{-\ri \rho_L}\frac{1+\mu_L^2}{2\mu_L}z_L^*\right)\oplus  \ri\left(\frac{1-\mu_R^2}{2\mu_R}z_R+e^{\ri \rho_R}\frac{1+\mu_R^2}{2\mu_R}z_R^*\right). \label{1012}\ee
 Choosing $\F=(\K\oplus\K)^{\rm diag}$, performing the isotropic gauging following the recipe of Section 3.3 and considering the case $\omega=0$, we arrive at the
$\sigma$-model action \eqref{999} with 
  \be \alpha=\frac{ \mu_R- \tan{\left(\frac{\rho_R}{2}\right)}}{ \mu_R +\tan{\left(\frac{\rho_R}{2}\right)} }.\frac{ \mu_L- \tan{\left(\frac{\rho_L}{2}\right)}}{ \mu_L +\tan{\left(\frac{\rho_L}{2}\right)} }.  \label{1011}\ee
The subspaces $\hat E_\pm$ needed for the RG calculations are nothing but the subspaces
$V^\perp_\pm$ identified explicitely in Eq.\eqref{743}
   \be \hat E_\pm =\left\{e^{\ri\left(\mp m_L-\jp\rho_L\right)}\frac{\sin{\left(\pm m_R+\frac{\rho_R}{2}\right)}}{\sin{(\rho_R)}}J_\pm\oplus
  e^{\ri\left(\mp m_R+\jp\rho_R\right)}\frac{\sin{\left(\mp m_L+\frac{\rho_L}{2}\right)}}{\sin{(\rho_L)}}J_\pm,\quad J_\pm\in\K\right\} \label{1015}\ee
  Here
\be \mu_a=\tan{(m_a)},\quad a=L,R.\label{932}\ee
As in the previous subsection, it is convenient to introduce certain  bijective maps $q_\pm:\K\to\hat E_\pm$. The following choice is the most convenient one
\be q_+(x)=\left(\cot{\left(m_L+\jp\rho_L\right)}-\ri \right)x  \oplus \left(\cot{\left(-m_R+\jp\rho_R\right)}+\ri \right)\frac{\sin{(\rho_R)}}{\sin{(\rho_L)}}\al x.\ee
\be q_-(x)=\left(\cot{\left(m_L-\jp\rho_L\right)}+\ri \right) \frac{\sin{(\rho_L)}}{\sin{(\rho_R)}}\al  x  \oplus \left(\cot{\left(-m_R-\jp\rho_R\right)}-\ri  \right)x.\ee

  Now we pick $x_\pm,u\in\K$, we evoke the definitions \eqref{766},\eqref{767} of the quantities $\al$, $\alpha_\pm$, $\beta$, $S_\pm^{L,R}$ and we  calculate the matrix elements of the flow operator $\hat \S_+$:
 \be\begin{aligned}  &\hat\P_+\ad_{q_+(x_+)}\hat\P_+^\perp\ad_{q_-(x_-)}\hat \P_+q_+(u)
=\\&=\frac{1}{\sin{\rho_L}}\hat\P_+\ad_{q_+(x_+)} \left(\frac{-\alpha^2\beta}{\sin{\rho_R}(S_-^LS_+^R)^2}([x_-,u]\oplus[x_-,u])+\frac{\alpha\alpha_-}{ S_-^LS_+^R}q_-[x_-,u]\right)=\\
 &=-\frac{\alpha^2(\beta+\alpha_+\alpha_-)}{\sin{\rho_L} \sin{\rho_R}(S_-^LS_+^R)^2} q_+([x_+,[x_-,u]])=\\&-\al\left(\cot{\Bigl(m_L+\jp\rho_L\Bigr)}-\ri  -\frac{\alpha\alpha_-}{\sin{\rho_L}S_-^LS_+^R}\right)\left(\frac{\sin{\rho_L}}{\sin{\rho_R}}\Bigl(\cot{\Bigl(-m_L+\jp\rho_L\Bigr)}-\ri \Bigr)-\frac{\alpha_+}{\sin{\rho_R}S_-^LS_+^R}\right)q_+([x_+,[x_-,u]])=\\&=\frac{\al\sin{\rho_L}}{\sin{\rho_R}}\left\vert \cot{\Bigl(m_L+\jp\rho_L\Bigr)}-\ri  -\frac{\al\alpha_-}{\sin{\rho_L}S_-^LS_+^R}\right\vert^2q_+([x_+,[x_-,u]])=\\&=
\frac{\al\Bigl((\al-1)^2+4\al\sin^2{\frac{\rho_L-\rho_R}{2}}\Bigr)\Bigl((\al-1)^2+4\al\sin^2{\frac{\rho_L+\rho_R}{2}}\Bigr)}{\sin{\rho_L}\sin{\rho_R}(\al^2-1)^2}q_+([x_+,[x_-,u]]).
 \end{aligned}\label{949}\ee 
 In the course of this calculation, the following  trigonometric identities were particularly useful
 \be \cot{\Bigl(\mp m_L+\jp\rho_L\Bigr)\sin{\rho_L}=\frac{S^R_\mp}{S^R_\pm}+\cos{\rho_L}}, \qquad \frac{1}{S^{a}_+S^{a}_-}-\frac{S^{a}_+}{S^{a}_-}-\frac{S^{a}_-}{S^{a}_+}
 =2\cos{\rho_{a}}, \quad a=R,L.\label{1034}\ee
 
 We infer from \eqref{949} that
 \be \frac{\tr\left(\hat\P_+\ad_{q_+(x_+)}\hat\P_+^\perp\ad_{q_-(x_-)}\hat \P_+\right)}{\tr(x_+x_-)}=c_K\frac{\al\Bigl((\al-1)^2+4\al\sin^2{\frac{\rho_L-\rho_R}{2}}\Bigr)\Bigl((\al-1)^2+4\al\sin^2{\frac{\rho_L+\rho_R}{2}}\Bigr)}{\sin{\rho_L}\sin{\rho_R}(\al^2-1)^2},\ee
 therefore it holds for the RG variation
 $$\frac{ 
 \Bigl(q_-(x_-), \delta(q_+(x_+))\Bigr)_\D}{\tr(x_+x_-)}=\frac{ 
 \left(q_-(x_-), \hat\S_+(q_+(x_+))\right)_\D}{2\tr(x_+x_-)}\delta s=-\frac{\tr(\hat\P_+\ad_{q_+(x_+) }\hat\P_+^\perp\ad_{q_-(x_-) }\hat\P_+)}{2\tr(x_+x_-)}\delta s=$$\be =-c_K\frac{\al\Bigl((\al-1)^2+4\al\sin^2{\frac{\rho_L-\rho_R}{2}}\Bigr)\Bigl((\al-1)^2+4\al\sin^2{\frac{\rho_L+\rho_R}{2}}\Bigr)}{2\sin{\rho_L}\sin{\rho_R}(\al^2-1)^2}\delta s.\label{1042a}\ee
 Now we use the identities
   \be \cot{\Bigl(m_L+\jp\rho_L\Bigr)}\sin{\rho_L}
 +\alpha\cot{\Bigl(m_R-\jp\rho_R\Bigr)}\sin{\rho_R}=
 \cos{\rho_L}-\al\cos{\rho_R},\ee
 \be \al\cot{\Bigl(m_L-\jp\rho_L\Bigr)}\sin{\rho_L}
 + \cot{\Bigl(m_R+\jp\rho_R\Bigr)}\sin{\rho_R}=
 \cos{\rho_R}-\al\cos{\rho_L},\ee
 to write 
 \be q_+(x_+)=\left( \frac{\cos{\rho_L}-\al\cos{\rho_R}}{2\sin{\rho_L}} x_+-\ri x_+\oplus -\frac{\cos{\rho_L}-\al\cos{\rho_R}}{2\sin{\rho_L}} x_+ + \ri\frac{\sin{\rho_R}}{\sin{\rho_L}}\al x_+\right) +\phi_+(x_+),\label{1048}\ee
  \be q_-(x_-)=\left(\frac{ \cos{\rho_R}-\al\cos{\rho_L}}{2\sin{\rho_R}}x_-+\ri\frac{\sin{\rho_L}}{\sin{\rho_R}}\al x_-\oplus -\frac{ \cos{\rho_R}-\al\cos{\rho_L}}{2\sin{\rho_R}}x_--\ri x_-  \right) +\phi_-(x_-).\label{1052}\ee
 where  $\phi_\pm(x_\pm)$  are certain quantities  belonging to  $\F$ that we do not need to know explicitely for our purposes; we need however the obvious fact that   $\delta\phi_+(x_+)$  belongs to $\F$.
 
 Thanks to \eqref{1048}, we  obtain
$$ \frac{ 
 \Bigl(q_-(x_-), \delta q_+(x_+)\Bigr)_\D}{\tr(x_+x_-)}=
 \delta\al 
 \frac{ 
 \Bigl(q_-(x_-), -\jp \cos{\rho_R}\ \!x_+\oplus\jp \cos{\rho_R} \ \!x_+
 +\ri\sin{\rho_R} \ \!x_+\Bigr)_\D}{{\sin{\rho_L}}\tr(x_+x_-)}=$$
 {\footnotesize \be =\delta\al 
 \frac{ 
 \Bigl(\jp( \cos{\rho_R}-\al\cos{\rho_L})x_-+\ri\al\sin{\rho_L}x_-\oplus - \jp( \cos{\rho_R}-\al\cos{\rho_L})x_--\ri \sin{\rho_R} x_-, -\jp \cos{\rho_R}\ \!x_+\oplus\jp \cos{\rho_R} \ \!x_+
 +\ri\sin{\rho_R} \ \!x_+\Bigr)_\D}{\sin{\rho_R}\sin{\rho_L}\tr(x_+x_-)}\ee}
 therefore
 \be  \frac{ 
 \Bigl(q_-(x_-), \delta q_+(x_+)\Bigr)_\D}{\tr(x_+x_-)}=-\frac{4\kappa\delta\al}{\sin{\rho_R}\sin{\rho_L}}.\label{1066}\ee
 Putting together Eqs.\eqref{1042a} and \eqref{1066}, we obtain the RG flow of the parameter $\alpha$:
 \be\frac{d\al}{ds}=c_K\frac{\al\Bigl((\al-1)^2+4\al\sin^2{\frac{\rho_L-\rho_R}{2}}\Bigr)\Bigl((\al-1)^2+4\al\sin^2{\frac{\rho_L+\rho_R}{2}}\Bigr)}{8\kappa(\al^2-1)^2}.\label{1069}\ee
If we  use the alternative set of parameters given by (cf. Eq. \eqref{756})
    \be \ \ka=\ka,\quad \rho_L=2\ka b_L,\quad \rho_R=2\ka b_R, \quad \al=e^{-2\ka  a}, \label{1072}\ee
    the flow formula \eqref{1069} gets rewritten as 
    \be \frac{da}{ds}=-c_K\frac{\left(\{a\}_\ka^2
    +[b_L-b_R]_\ka^2\right)\left(\{a\}_\ka^2
    +[b_L+b_R]_\ka^2\right)}{16\{a\}_{2\ka}^2},\label{1073}\ee
    where we have borrowed the notation from the
    "$q$-deformed literature":
    \be \{x\}_\ka:=\frac{\sinh{(\ka x)}}{\ka},\qquad [x]_\ka:=\frac{\sin{(\ka x)}}{\ka}.\ee
    The flow formula \eqref{1073}  lends itself perfectly to the study of the limit $\kappa\to 0$ which was performed at the end of Section 4 to show that  the bi-Yang-Baxter deformation of the WZW model tends to the bi-Yang-Baxter deformation of the principal chiral model. Does the flow formula
    \eqref{1073} of the former deformation go in this limit to the flow formula \eqref{865} of the latter? Yes, it does because it obviously holds
    \be \lim_{\ka\to 0}\{x\}_\ka=x,\qquad \lim_{\ka\to 0}[x]_\ka=x.\ee
    
   There are two more special cases, where our flow formulae \eqref{1069} or \eqref{1073}  can be compared with  the results already obtained in the literature. First one corresponds to the single Yang-Baxter deformation where $\rho_L=0$ and $\rho_R\neq 0$. Upon the transformation \eqref{739}, an easy calculation shows that the flow \eqref{1069} matches exactly the flow of the YB-WZ model as obtained in Ref.\cite{DDST}. The second special case is the Lukyanov flow \cite{L12}
    which should coincide with our flow \eqref{1069} for the choice $K=SU(2)$. To verify this is technically more involved, because it is necessary  to introduce coordinates on the group manifold, and we devote an entire next subsection to this task.

\subsection{Comparison with the Lukyanov flow}

Lukyanov model is a non-linear $\sigma$-model living on the target of the group $SU(2)$. It was introduced in Ref.\cite{L12} and, in the case of the vanishing TsT parameter, its target space geometry is characterized by the following metric $G$ and the Kalb-Ramond field $B$:
$$ G=\frac{(\vk+P)(\vk+P^{-1})}{g^2}\frac{dz^2}{(1-z^2)(1-\varkappa^2z^2)}+\frac{1}{g^2}\frac{1}{1-\vk^2z^2 +\vk Q(1-z^2)}\times$$
\be \times\left[(1-\vk^2z^2+\vk P(1-z^2)  )dv^2+(1-\vk^2z^2+\vk P^{-1}(1-z^2)  )dw^2-2(1-\vk^2)z dw dv\right],\label{1100}\ee
\be B=\frac{1}{g^2} \sqrt{ \frac{(\vk+P)(\vk+P^{-1})}{  (\vk+Q)(\vk+Q^{-1})}}\frac{(1-\vk^2)z}{1-\vk^2z^2 +\vk Q(1-z^2)}dv\wedge dw.\label{1101}\ee
Here $z,v,w$ are appropriate coordinates on the group $SU(2)$ which will be specified in what follows and $\vk,P,Q,g$ are  real free parameters of the model restricted by Lukyanov to the values  $g>0$, $\vk\in]0,1[$, $P>0$  and $Q>0$ (actually, $P$ and $Q$ were respectively denoted in \cite{L12} as $p^2$ and $h^2/\vk$ and $B$ given by \eqref{1101} differs from Lukyanov's one by an inessential  total derivative). 
  
  The RG flow of the parameters was found in \cite{L12} and it is given by 
\be \dot P=0,\quad \dot Q=0, \quad \dot\vk=-\frac{g^2\vk(1-\vk^2)}{(\vk+P)(\vk+P^{-1})}, \quad  
 \dot g=\frac{g^3(1-\vk^2)^2}{4(\vk+P)^2(\vk+P^{-1})^2}\left(1-\frac{(\vk+P)(\vk+P^{-1})}{(\vk+Q)(\vk+Q^{-1})}\right).\label{1117}\ee
 
Our goal in this subsection is to compare the Lukyanov target space data \eqref{1100} and \eqref{1101} with those extracted from the general bi-YB-WZ action \eqref{999}  for the special case of the group $K=SU(2)$. We do find the perfect match both of the target space geometry and of the RG flow provided we carefully adjust  the ranges of the Lukyanov parameters and of the bi-YB-WZ ones. We make this adjusting at the very end of the present section.   

We now write
the bi-YB-WZ action
 \be S_{\rm bi-YB-WZ}(k)= \kappa I_{\rm WZ}(k)  + \kappa \int d\tau \oint \  \tr k^{-1}\partial_+k \frac{\al+e^{\rho_LR_k}e^{\rho_RR}}{\al-e^{\rho_LR_k}e^{\rho_RR}}  k^{-1}\partial_-k\label{1124f}\ee
 in the standard coordinates on the $SU(2)$ group manifold 
 \be k = \bm e^{\ri\phi}&0\\ 0&e^{-\ri\phi}\em  \bm \cos{\theta}&\ri\sin{\theta}\\\ri\sin{\theta} &\cos{\theta}\em\bm e^{\ri\psi}&0\\ 0&e^{-\ri\psi}\em.\ee
We find first
 \be k^{-1}\d_\pm k=\bm \ri(\d_\pm\psi+\cos{(2\theta)}\d_\pm\phi)&e^{-2\ri\psi}(\ri\d_\pm\theta-\sin{(2\theta)}\d_\pm\phi)\\e^{2\ri\psi}(\ri\d_\pm\theta+\sin{(2\theta)}\d_\pm\phi)&-\ri(\d_\pm\psi+\cos{(2\theta)}\d_\pm\phi)\em.\ee
The Yang-Baxter operator $R$ acts on the elements of the Lie algebra $\K=su(2)$ as
\be R\bm \ri z& y+\ri x\\ -y+\ri x& -\ri z\em = 
\bm 0& x-\ri y\\ -x-\ri y& 0\em =\jp\left[\bm -\ri& 0\\0& \ri\em, \bm \ri z& y+\ri x\\ -y+\ri x& -\ri z\em\right],\quad x,y,z\in\br,\ee
we thus infer that
\be e^{\rho_RR}\bm \ri z& y+\ri x\\ -y+\ri x& -\ri z\em =\bm e^{-\ri\rjp}&0\\0&e^{\ri\rjp}\em \bm \ri z& y+\ri x\\ -y+\ri x& -\ri z\em \bm e^{\ri\rjp}&0\\0&e^{-\ri\rjp}\em.\ee
Define  elements $T\in SU(2)$ and $\chi_\pm\in su(2)$ as follows
\be  T:=  \bm \cos{\theta}&-\ri\sin{\theta}\\-\ri\sin{\theta} &\cos{\theta}\em \bm e^{-\ri\ljp}&0\\0&e^{\ri\ljp}\em \bm \cos{\theta}&\ri\sin{\theta}\\\ri\sin{\theta} &\cos{\theta}\em \bm e^{-\ri\rjp}&0\\0&e^{\ri\rjp}\em,\ee
 \be \chi_\pm :=\bm \ri(\d_\pm\psi+\cos{(2\theta)}\d_\pm\phi)& \ri\d_\pm\theta-\sin{(2\theta)}\d_\pm\phi\\ \ri\d_\pm\theta+\sin{(2\theta)}\d_\pm\phi&-\ri(\d_\pm\psi+\cos{(2\theta)}\d_\pm\phi)\em.\ee
 Then the action \eqref{1124f} can be rewritten as
  \be S_{\rm bi-YB-WZ}(k)= \kappa I_{\rm WZ}(k)  + \kappa \int d\tau \oint d\sigma\  \tr \chi_+ \frac{\al+{\rm Ad}_T}{\al-{\rm Ad}_T}  \chi_-.\label{1134}\ee
  We associate to the element $T\in SU(2)$ an angle  $\xi\in[0,\pi]$  and an element $t\in su(2)$  as follows
  \be \xi=\arccos{\left(\jp\tr T\right)}, \quad t=\frac{T-(\cos{\xi})\1}{\sin{\xi}}.\ee
  Here $\1$ is the unit matrix and we note that 
  the singular values $\xi=0,\pi$ are avoided  because of the non-vanishing parameters $\rho_L,\rho_R$. Note also that it holds
  \be \cos{\xi}=\col\cor-\sl\sr\cos{2\theta}.\label{1150}\ee
  
  As $T$ varies with $\theta$, the element $t$ sweeps a hyperplane in $su(2)$; indeed, we can check easily that it holds
  \be \tr(rt)=0\ee
where $r$ is the following $\theta$-independent element of $su(2)$:
\be r= \bm 0&\ri e^{\ri\rjp}\\\ri e^{-\ri\rjp}&0\em.\ee
The crucial fact needed to evaluate the
action \eqref{1134} is the validity of the following identities
\be Ad_T t=t,\quad Ad_T(rt\pm \ri r)=e^{\mp\ri2\xi}(rt\pm \ri r),\label{1146}\ee
\be \tr t^2=-2, \quad \tr r^2= -2, \quad \tr (rt)^2=-2,\quad \tr tr=0, \quad \tr r(rt)=0, \quad \tr t(rt)=0.\label{1148}\ee
We then infer from \eqref{1146} and \eqref{1148} 
$$ \chi_\pm = -\jp\tr(t\chi_\pm)t-\jp\tr(r\chi_\pm)r
-\jp\tr(rt\chi_\pm)rt=$$\be = -\jp\tr(t\chi_\pm)t-\js\tr((rt+\ri r)\chi_\pm)(rt-\ri r)
-\js\tr((rt-\ri r)\chi_\pm)(rt+\ri r)\ee
and, subsequently,
\be\frac{\al+{\rm Ad}_T}{\al-{\rm Ad}_T}  \chi_-= - \jp\frac{\al+ 1}{\al-1}  \tr (t\chi_-)t- \js\frac{\al+e^{-\ri 2\xi}}{\al-e^{-\ri 2\xi}}  \tr( (rt-\ri r)\chi_-)(rt+\ri r)  -\js\frac{\al+e^{\ri 2\xi}}{\al-e^{\ri 2\xi}} \tr((rt+\ri r)\chi_-)(rt-\ri r) .\ee
   Putting all together, we find
 $$ S_{\rm bi-YB-WZ}(k)= \kappa I_{\rm WZ}(k)-\frac{\kappa}{2} \frac{\al+1}{\al-1}\int d\tau \oint d\sigma \tr (t\chi_+)\tr (t\chi_-)$$$$ -\frac{\ka}{4} \int d\tau \oint d\sigma \biggl(\frac{\al+e^{\ri 2\xi}}{\al-e^{\ri 2\xi}}\tr((rt+\ri r)\chi_-)  \tr((rt-\ri r)\chi_+) +\frac{\al+e^{-\ri 2\xi}}{\al-e^{-\ri 2\xi}}\tr((rt-\ri r)\chi_-)  \tr((rt+\ri r)\chi_+) \biggr)=$$
 $$ =\kappa I_{\rm WZ}(k)-\frac{\kappa}{2} \int d\tau \oint d\sigma \biggl( \frac{\al+1}{\al-1}\tr (t\chi_+)\tr (t\chi_-)  + \frac{\al^2-1}{\al^2+1-2\al\cos{2\xi}}  \Bigl(\tr(r\chi_+)\tr (r\chi_-)+\tr(rt\chi_+)\tr (rt\chi_-)\Bigr)\biggr)$$
 \be +\ka\int d\tau \oint d\sigma  \frac{\al\sin{2\xi}}{\al^2+1-2\al\cos{2\xi}}  \Bigl(\tr(rt\chi_+)\tr(r\chi_-)-\tr(r\chi_+)\tr(rt\chi_-)\Bigr).
   \label{1157}\ee
   We find easily
    \be \tr(r\chi_\pm)=-2\cor\d_\pm\theta-2\sr\sin{2\theta}\ \!\d_\pm\phi,\label{1161}\ee
    \be \tr(rt\chi_\pm)=\frac{2}{\sin{\xi}}
    \left(\left(\col-\cor\cos{\xi}\right)\d_\pm\theta+\sl\sin{2\theta}\ \!\d_\pm\psi-\sr\sin{2\theta}\cos{\xi}\ \!\d_\pm\phi\right).\label{1162}\ee
    $$\tr(t\chi_\pm)=-\frac{2\sin{2\theta}}{\sin{\xi}} \sin{\ljp}\sr\d_\pm\theta+$$\be +
  \frac{2}{\sin{\xi}} \left(\sl\cor+\sr\col\cos{2\theta}\right)\d_\pm\phi + \frac{2}{\sin{\xi}} \left(\sr\col+\sl\cor\cos{2\theta}\right)\d_\pm\psi ,\label{1171}\ee
  Combining Eqs.\eqref{1157}, \eqref{1161}, \eqref{1162}  and \eqref{1171}, we
 then find the following background metric $ds^2$ and the Kalb-Ramond field $B$
  \be \frac{1}{2\ka}\frac{1-\al}{1+\al} ds^2=   
  d\theta^2+d\psi^2+d\phi^2 + 2\cos{2\theta}d\psi d\phi -\frac{4\al\sin^{2}{2\theta} }{(\al-1)^2+4\al\sin^2{\xi}}\left( d\tilde\psi^2+ d\tilde\phi^2 -2\cos{\xi}   d\tilde\psi d\tilde\phi\right),\label{1180}\ee
  \be B=\ka \cos{2\theta}d\psi\wedge d\phi+ \frac{4\ka\al\sin^2{2\theta}}{(\al-1)^2+4\al\sin^2{\xi}}\cos{\xi}d\tilde\psi\wedge d\tilde\phi.\label{1181}\ee
where  
\be d\tilde\psi=\sl\left(d\psi+\frac{\cot{\ljp}}{\sin{2\theta}}d\theta\right), \quad  d\tilde\phi=\sr\left(d\phi+\frac{\cot{\rjp}}{\sin{2\theta}}d\theta\right).\label{1182}\ee

Now we trade the parameters $\ka$, $\al$, $\rho_L$ and $\rho_R$ for the parameters $g$, $\vk$, $P$ and $Q$ as follows\footnote{  For completeness, we list also the reciprocal transformation of the parameters:
$$ \ka=\frac{1}{2g^2}
 \sqrt{\frac{(\vk+P)(\vk+P^{-1})}{(\vk+Q)(\vk+Q^{-1})}}, \  \tan^2{\frac{\rho_L}{2}}= -QP, \  \tan^2{\frac{\rho_R}{2}}= -QP^{-1},\  \al =\frac{\sqrt{(\vk+Q)(\vk+Q^{-1})}-\sqrt{(\vk+P)(\vk+P^{-1})}}{\sqrt{(\vk+Q)(\vk+Q^{-1})}+\sqrt{(\vk+P)(\vk+P^{-1})}}$$}
 \be g^2=\frac{1}{2k}.\frac{1-\al}{1+\al},\quad  P=-\frac{\tan{\frac{\rho_L}{2}}}{\tan{\frac{\rho_R}{2}}}, \quad Q= \tan{\frac{\rho_L}{2}}\tan{\frac{\rho_R}{2}};\label{1121}\ee
 \be
 \vk=\frac{\sqrt{(\al-1)^2+4\al\sin^2{\frac{\rho_L+\rho_R}{2}}}-\sqrt{(\al-1)^2+4\al\sin^2{\frac{\rho_L-\rho_R}{2}}}}{\sqrt{(\al-1)^2+4\al\sin^2{\frac{\rho_L+\rho_R}{2}}}+\sqrt{(\al-1)^2+4\al\sin^2{\frac{\rho_L-\rho_R}{2}}}}\label{1122},\ee
and, at the same time, we change the coordinates on the target $K=SU(2)$ according to the formulas 
$$ z=\frac{
\cos{2\theta}+\vk}{1+\vk\cos{2\theta}},$$ 
\be   v=\psi +\frac{1}{4}\tar\ln{\left(1+\vk^2+2\vk\cos{2\theta}\right)},   \quad w=-\phi -\frac{1}{4}\tal  \ln{ \left(1+\vk^2+2\vk\cos{2\theta}\right)}.\label{1194}\ee
With these changements of the parameters and of the coordinates on the target, the metric \eqref{1180}  becomes exactly the Lukyanov metric \eqref{1100} and the Kalb-Ramond field \eqref{1181} becomes, up to a total derivative,
 the Kalb-Ramond field   \eqref{1101}. The calculation proving this fact is tedious but straightforward and it is   simplified by the repeated use of the following formula (valid for  $\al\neq 0$, $\vk\neq 0)$
 \be \frac{(1-\vk)^2}{4\vk}=\frac{ \frac{(1-\al)^2}{4\al}+\sin^2{\frac{\rho_L-\rho_R}{2}}}{\sin{\rho_L}\sin{\rho_R}}.\ee
Moreover, it can be checked directly that the transformation of the parameters \eqref{1121}, \eqref{1122} transforms the bi-YB-WZ flow \eqref{1069} into the Lukyanov flow   \eqref{1117} 
 (note that $c_K=4$ for $su(2)$, the parameters $\ka$, $\rho_L$, $\rho_R$ do not flow and the Lukyanov RG time of Ref.\cite{L12} runs in the opposite direction with respects to the conventions of Sections 6.2 and 6.3  commonly used in the Poisson-Lie literature).
 
 It remains to discuss the issue of  the possible ranges of the Lukyanov
 parameters $g$, $\vk$, $P$ and $Q$ and of the bi-YB-WZ parameters
 $\al$, $\ka$, $\rho_L$ and $\rho_R$. In the Lukyanov paper, all parameters are positive or non-negative, more precisely, he considered the case  $g>0$, $P>0$, $Q>0$ and $0\leq \vk< 1$. Looking at Eq. \eqref{1121}, we observe that this choice is out of reach\footnote{However, if we permitted imaginary values of $\rho_L$ and $\rho_R$ then by an appropriate analytic continuations of the  target space coordinates  we would reach the Lukyanov model within the range of the parameters that he considered.} of the bi-YB-WZ model where $P$ has necessarily the opposite sign with respect to $Q$.  On the other hand, the Lukyanov geometry \eqref{1100} and \eqref{1101}
 makes perfect sense (in particular the  metric remains positive definite) for  a wider range of the  parameters than  he considered in Ref.\cite{L12}. This extended consistent range is given by \be g>0, \quad \vk\in]-1,1[, \quad  P\neq 0, \quad Q\neq 0, \quad 1+\vk P^{\pm 1} >0,  \quad  1+\vk Q^{\pm 1} >0.  \label{1202}\ee
The bi-YB-WZ model for $SU(2)$ turns out to match the extended range Lukyanov model without any need of analytical continuation. Indeed, for all admissible values of the bi-YB-WZ parameters, i.e. $\ka\in\mathbb N$, $-1<\al<1$, $0<\vert\rho_L\vert<\pi$ and $0<\vert\rho_R\vert<\pi$,   a careful analysis shows that the Lukyanov parameters  $g,\vk,P,Q$ given by the ranges of the functions \eqref{1121}, \eqref{1122} always respect the extended range conditions \eqref{1202}.

  \section{Outlook}
  \setcounter{equation}{0}
 The present work solves two from the open problems listed in the outlook of Ref.\cite{DHKM}, namely, it provides the $\E$-model formulation of
 the DHKM model and, also, it  settles the issue of the renormalizability.  We believe, that the $\E$-model insight should be helpful also for tackling the remaining open question
 from the list, which  is the status of the Hamiltonian integrability of the model. 
 
 The dressing coset structure of the DHKM model indicates the occurrence of
 a rich T-duality story which should go  well beyond the simple T-duality corresponding to the changing of the TsT parameters. 
 In particular, the example of the Poisson-Lie T-duality \eqref{741} occurring in the YB-WZ model should  generalize to the bi-YB-WZ context.  How it happens precisely remains to be worked out.

 \vskip2pc

  \noindent{\bf Acknowledgement}: I thank to G. Kotousov for his highly valued  computer work help and  to V. Bazhanov, G. Kotousov and S. Lukyanov for inspiring discussions. I  gratefully acknowledge the  support from the Simons Center for Geometry and Physics, Stony Brook University at which some of the research for this paper was performed.


\begin{thebibliography}{99}
 
 \bibitem{AOSSY}{T. Araujo, E.  \'O Colg\'ain, Y. Sakatani, M.M. Sheikh-Jabbari and H.Yavartanoo, {\it  Holographic integration of $T\bar T$ and  $J\bar T$ via $O(d,d)$},  JHEP {\bf 1903} (2019) 168, arXiv:1811.03050 [hep-th]}

\bibitem{ABR}{C. Ahn, J. Balog and F. Ravanini, {\it Nonlinear integral equations for the sausage model},
  J.Phys. {\bf A50} (2017) no.31, 314005 }
\bibitem{AABL93}{E. Alvarez, L. Alvarez-Gaum\'e, J. Barb\'on and Y. Lozano, {\it Some global aspects of duality in string theory}, Nucl. Phys. {\bf B415} (1994) 71, hep-th/9309039} 

\bibitem{AHPT}{C. Appadu, T.J. Hollowood, D. Price and  D.C. Thompson, {\it Quantum Anisotropic Sigma and Lambda Models as Spin Chains},
 J.Phys. {\bf A51} (2018) no.40, 405401, 
  arXiv:1802.06016 [hep-th]}
  
   \bibitem{ABF}{ G. Arutyunov, R. Borsato and S. Frolov, 
{\it $S$-matrix for strings on $\eta$-deformed $AdS_5 \times S^5$},	JHEP (2014) 002,  arXiv:1312.3542 [hep-th]}

\bibitem{Mad93}{J. Balog, P. Forg\'acs, Z. Horv\'ath and L. Palla, {\it A new family of $SU(2)$ symmetric  integrable $\sigma$-models}, Phys. Lett. {\bf B324} (1994) 403, hep-th/9307030}

\bibitem{BM}{I. Bakhmatov and E. Musaev, {\it Classical Yang-Baxter equation from $\beta$-supergravity},
  JHEP {\bf 1901} (2019) 140, 
  arXiv:1811.09056 [hep-th]}
  
  \bibitem{BKL}{
V.V. Bazhanov, G.A. Kotousov and S.L. Lukyanov, {\it Winding vacuum energies in a deformed O(4) sigma model},  
  Nucl.Phys. {\bf B889} (2014) 817-826,
  arXiv:1409.0449 [hep-th]; {\it 
Quantum transfer-matrices for the sausage model},
 JHEP {\bf 1801} (2018) 021, 
  arXiv:1706.09941 [hep-th]; {\it On the Yang-Baxter Poisson algebra in non-ultralocal integrable systems},  Nucl. Phys. {\bf B934} (2018) 529,  arXiv:1805.07417 [hep-th]}
 


 

  \bibitem{BTW} {  R. Borsato, A.A. Tseytlin and L. Wulff, {\it 
 Supergravity background of $\lambda$-deformed model for $AdS2\times S2$ supercoset},
 Nucl.Phys. {\bf B905} (2016) 264-292,
  arXiv:1601.08192 [hep-th]}
  
  \bibitem{By}{D. Bykov, {\it Flag manifold sigma-models: the $1/N$-expansion and the anomaly two-form}, Nucl. Phys. {\bf B941} (2019) 316, arXiv:1901.02861 [hep-th];  {\it  Complex structure-induced deformations of sigma models},
 JHEP {\bf 1703} (2017) 130,  
  arXiv:1611.07116 [hep-th]}


  
    \bibitem{CM06} {A. Cabrera and  H. Montani, {\it Hamiltonian loop group actions and T-duality for group manifolds}, 
 J.Geom.Phys. {\bf 56} (2006) 1116-1143,
 hep-th/0412289}
 
  \bibitem{C81} {I. V. Cherednik; {\it Relativistically invariant quasiclassical limits of integrable two-dimensional quantum models},
Theor. Math. Phys. {\bf 47} (1981) 422}

\bibitem{CL}{Y. Chervonyi and O. Lunin, {\it Generalized $\lambda$-deformations of $AdS_p\times S^p$}, Nucl. Phys. {\bf B913} (2016) 912, arXiv: 1608.06641 [hep-th]}

\bibitem{DHKLM}{F. Delduc, B.Hoare, T. Kameyama, S. Lacroix and  M. Magro, {\it Three-parameter integrable deformation of $\mathbb Z_4$ permutation supercosets}, 
  JHEP {\bf 1901} (2019) 109,  arXiv:1811.00453 [hep-th]}


  \bibitem{DHKM}{F. Delduc, B. Hoare, T. Kameyama and  M. Magro,
 {\it Combining the bi-Yang-Baxter deformation, the Wess-Zumino term and TsT transformations in one integrable $\sigma$-model}, JHEP {\bf 1710} (2017) 212, arXiv:1707.08371 [hep-th]}
 
 
 
\bibitem{DMV13}{ F. Delduc, M. Magro and B. Vicedo, {\it  On classical q-deformations of integrable sigma-models},  JHEP{\bf 11} (2013) 192,
 arXiv:1308.3581 [hep-th]; { \it  An integrable deformation of the $AdS_5 \times  S^5$  superstring action}, Phys. Rev. Lett. {\bf 112} (2014) 051601, arXiv:1309.5850 [hep-th]}

   \bibitem{DMV15}{F. Delduc, M. Magro and  B. Vicedo, {\it  Integrable double deformation of the principal chiral model}, Nucl. Phys. {\bf B891} (2015) 312-321, arXiv:1410.8066 [hep-th]}
   
   \bibitem{DLMV}{F. Delduc, S. Lacroix, M. Magro and B. Vicedo, {\it Integrable
  coupled sigma-models},
 arXiv:1811.12316 [hep-th]; {\it On the Hamiltonian integrability of the bi-Yang-Baxter $\sigma$-model},
   JHEP {\bf 1603} (2016) 104,
 arXiv:1512.02462 [hep-th]}
 
 \bibitem{DHT}{S.~Demulder, F.~Hassler and D.~C.~Thompson, {\it Doubled aspects of generalised dualities and integrable deformations},  JHEP {\bf 1902} (2019) 189,
  arXiv:1810.11446 [hep-th], {\it An invitation to Poisson-Lie T-duality in double field theory and its applications}, arXiv:1904.09992 [hep-th]}
  
  \bibitem{DST}{S.~Demulder, K.~Sfetsos and D.~C.~Thompson,
  {\it Integrable $\lambda$-deformations: Squashing Coset CFTs and $AdS_5\times S^5$},
  JHEP {\bf 1507} (2015) 019,
 arXiv:1504.02781 [hep-th]}
 
 \bibitem{DDST}{S. Demulder, S. Driezen, A. Sevrin and D. Thompson, {\it Classical and Quantum Aspects of Yang-Baxter Wess-Zumino Models}, 
  JHEP {\bf 1803} (2018) 041, 
  arXiv:1711.00084 [hep-th]}
  
  \bibitem{DZ}{H. Dlamini and K. Zoubos, {\it Marginal deformations and quasi-Hopf algebras}, arXiv:1902.08166 [hep-th]}
 
 \bibitem{Fa}{V.A. Fateev, {\it The sigma model (dual) representation for a two-parameter family of integrable quantum field theories}, Nucl. Phys. {\bf B473} (1996) 509}
 
 \bibitem{Fa19}{V.A. Fateev, {\it Classical and quantum integrable sigma models. Ricci flow, "nice duality" and perturbed rational conformal field theories}, arXiv:1902.02811 [hep-th]}
 
 \bibitem{FaLi}{V.A. Fateev and A.V. Litvinov, {\it Integrability, Duality and Sigma Models},   JHEP {\bf 1811} (2018) 204, arXiv:1804.03399 [hep-th] }

\bibitem{FOZ}{
 V.A. Fateev, E. Onofri and A.B. Zamolodchikov, {\it The Sausage model (integrable deformations of $O(3)$ $\sigma$-model)}, Nucl. Phys. {\bf B406} (1993) 521}
 
\bibitem{TSSY}{
J. J. Fernandez-Melgarejo, J. Sakamoto, Y. Sakatani  and K. Yoshida, {\it  T-folds from Yang-Baxter deformations},
 JHEP {\bf 1712} (2017) 108, 
  arXiv:1710.06849 [hep-th]}
  
  \bibitem{FR}{S. F\"orste and D. Roggenkamp, {\it Current-current deformations of conformal field theories, and WZW models},  JHEP {\bf 0305} (2003) 071, hep-th/0304234} 
 
 \bibitem{F}{S. Frolov, {\it Lax Pair for strings in Lunin-Maldacena background}, JHEP {\bf 0505} (2005) 069,  hep-th/0503201}
 \bibitem{GS}{ G.~Georgiou and K.~Sfetsos,
  {\it A new class of integrable deformations of CFTs},
  JHEP {\bf 1703} (2017) 083,
  arXiv:1612.05012 [hep-th]; {\it 
  The most general $\lambda$-deformation of CFTs and integrability}, JHEP {\bf 1903}  (2019) 094,
  arXiv:1812.04033 [hep-th]}
  
\bibitem{GSS}{
  G.~Georgiou, K.~Sfetsos and K.~Siampos,
  {\it Double and cyclic $\lambda$-deformations and their canonical equivalents},
  Phys.\ Lett.\ B {\bf 771} (2017) 576, 
  arXiv:1704.07834 [hep-th]}
  
  \bibitem{GPZ}{D. Giataganas, L. Pando Zayas and K. Zoubos, {\it On marginal deformations and non-integrability}, JHEP {\bf 1401} (2014) 129, arXiv:1311.3241 [hep-th]}
 
  \bibitem{Ha} {F. Hassler, {\it Poisson-Lie T-Duality in Double Field Theory}, arXiv:1707.08624 [hep-th]}
 




  \bibitem{GR93}{A. Giveon and M. Ro\v cek, {\it On nonAbelian duality}, Nucl.Phys. {\bf B421} (1994) 173-190, hep-th/9308154}
  \bibitem{Hew96}{S.F. Hewson, {\it 
The Non-Abelian target space duals of Taub - NUT space},
  Class.Quant.Grav. {\bf 13} (1996) 1739-1750,
 hep-th/9510092}
 
  \bibitem{H}{B.~Hoare,
  {\it Towards a two-parameter q-deformation of AdS$_3 \times S^3 \times M^4$ superstrings},
  Nucl.\ Phys.\ {\bf B891} (2015) 259,
  arXiv:1411.1266 [hep-th]}



 
  \bibitem{BRT}{B.~Hoare, R.~Roiban and A.~A.~Tseytlin,
 {\it On deformations of $AdS_n\times S^n$ supercosets},
 JHEP {\bf 1406} (2014) 002,
 arXiv:1403.5517 [hep-th]}
 
     \bibitem{HS}{B. Hoare and F. Seibold, {\it	
Poisson-Lie duals of the $\eta$-deformed $AdS2\times S2\times T6$ superstring},
 JHEP {\bf 1808} (2018) 107, arXiv:1807.04608 [hep-th]}
 
 \bibitem{HT} {B. Hoare and A.A. Tseytlin, {\it On integrable deformations of superstring sigma models related to} $AdS_n\times S^n$ {supercosets}, Nucl. Phys. {\bf B897} (2015) 448, arXiv:1504.07213[hep-th]}
 \bibitem{KY}  {I. Kawaguchi and K. Yoshida,
 {\it Hybrid classical integrability in squashed sigma models}, Phys.Lett. {\bf B705} (2011) 251, arXiv:1107.3662  [hep-th];
  {\it A deformation of quantum affine algebra in squashed Wess-Zumino-Novikov-Witten models},
  J.\ Math.\ Phys.\  {\bf 55} (2014) 062302,
  arXiv:1311.4696 [hep-th]}
\bibitem{KMY11}{I. Kawaguchi, T. Matsumoto and K. Yoshida, {\it  On the classical equivalence of monodromy matrices in squashed sigma model}, JHEP {\bf 1206} (2012) 082, arXiv:1203.3400 [hep-th]}
 \bibitem{KOY11} {I.~Kawaguchi, D.~Orlando and K.~Yoshida,
  {\it Yangian symmetry in deformed WZNW models on squashed spheres},
  Phys.\ Lett.\ B {\bf 701} (2011) 475,
  arXiv:1104.0738 [hep-th]}

 
  \bibitem{K18}{C. Klim\v c\'\i k, {\it  
Affine Poisson and affine quasi-Poisson T-duality}, 
  Nucl.Phys. {\bf B939} (2019) 191-232, arXiv:1809.01614 [hep-th]}
  
 
 
\bibitem{K}{C. Klim\v c\'\i k, {\it  Yang-Baxter $\sigma$-model and dS/AdS T-duality}, JHEP {\bf 0212} (2002) 051, hep-th/0210095; {\it Integrability of the Yang-Baxter $\sigma$-model}, J. Math. Phys. {\bf 50} (2009) 043508, arXiv:0802.3518 [hep-th]; {\it Integrability of the bi-Yang-Baxter $\sigma$-model}, Lett. Math. Phys. {\bf 104} (2014) 1095, arXiv:1402.2105 [math-ph]}
 
 \bibitem{K15}
  {C.~Klim\v c\'\i k,
  {\it $\eta$ and $\lambda$ deformations as ${\mathcal E}$-models}, 
  Nucl.\ Phys. {\bf B900} (2015) 259, 
  arXiv:1508.05832 [hep-th]; {\it Poisson-Lie T-duals 
    of the bi-Yang-Baxter models}, Phys. Lett. {\bf B760} (2016)
    345-349, arXiv:1606.03016 [hep-th]}
  
  \bibitem{K17}{C. Klim\v c\'\i k, {\it 
Yang-Baxter $\sigma$-model with WZNW term as $\E$-model}, 
 Phys.Lett. {\bf B772} (2017) 725-730,  arXiv:1706.08912 [hep-th]} 

 

  
  \bibitem{KP99}{C. Klim\v c\'\i k and S. Parkhomenko, {\it Supersymmetric gauged WZNW models as dressing cosets}, 
 Phys.Lett. {\bf B463} (1999) 195-200,
  hep-th/9906163}
  
       \bibitem{KS95}{C. Klim\v c\'\i k and P. \v Severa, {\it Dual non-Abelian duality and the Drinfeld double},
Phys. Lett. {\bf B351}
(1995) 455-462, hep-th/9502122; C. Klim\v c\'\i k, {\it Poisson-Lie $T$-duality},
Nucl. Phys. (Proc. Suppl.) {\bf 
B46} (1996) 116-121, hep-th/9509095; P. \v Severa, 
{\it Minim\'alne plochy a dualita}, Diploma thesis, Prague University, 1995,  in Slovak}

\bibitem{KS96a}{C. Klim\v c\'\i k and P. \v Severa, {\it  Poisson-Lie T-duality and loop groups of Drinfeld doubles},  Phys. Lett. {\bf B372} (1996), 65-71,  hep-th/9512040}
 
\bibitem{KSopen}{C. Klim\v c\'\i k and P. \v Severa, {\it  Poisson-Lie T-duality: Open strings and D-branes}, Phys. Lett. {\bf B376} (1996) 82-89, hep-th/9512124}
  
     \bibitem{KS96b}{C. Klim\v c\'\i k and P. \v Severa, {\it Dressing cosets},
Phys. Lett. {\bf B381}
(1996) 56-61, hep-th/9602162}
  


 \bibitem{KS97}{C. Klim\v c\'\i k and P. \v Severa, {\it Non-Abelian momentum-winding exchange}, Phys.Lett. {\bf B383} (1996) 281-286, hep-th/9605212;  {\it Open strings and $D$-branes in WZNW model}, Nucl.Phys. {\bf B488} (1997) 653, hep-th/9609112}


\bibitem{KT94}{C. Klim\v c\'\i k and A. A. Tseytlin, {\it Exact four-dimensional string solutions and Toda like sigma models from 'null gauged' WZNW theories},
  Nucl.Phys. {\bf B424} (1994) 71-96,
  hep-th/9402120}
  
  \bibitem{Kot} {G.A. Kotousov, personal communication}
  
  \bibitem{KY2}{H. Kyono and K. Yoshida, {\it Supercoset construction of Yang-Baxter deformed $AdS_5\times S^5$ background}, {\bf PTEP} 2016 (2016) no.8, 083B03, arXiv:1605.02519 [hep-th]}
  
  \bibitem{Lit}{A.V. Litvinov, {\it Integrable $gl(n\vert n)$ Toda field theory and its sigma-model dual},
  arXiv:1901.04799 [hep-th] }

\bibitem{LS}{A.V. Litvinov and L.A. Spodyneiko, {\it On dual description of the deformed $O(N)$ sigma model},
   JHEP {\bf 1811} (2018) 139, arXiv:1804.07084 [hep-th]}

  \bibitem{L12} {
 S. L. Lukyanov, {\it The integrable harmonic map problem versus Ricci flow}, Nucl. Phys. {\bf B865} (2012) 308, arXiv:1205.3201  [hep-th]}

\bibitem{LM}{O. Lunin and J. Maldacena, {\it Deforming field theories with $U(1)\times U(1)$ global symmetry and their gravity duals}, JHEP {\bf 0505} (2005) 033, hep-th/0502086}

\bibitem{LT}{O. Lunin and W. Tian, {\it Analytical structure of the generalized $\lambda$-deformation}, Nucl. Phys. {\bf B929} (2018) 330,  arXiv: 1711.02735 [hep-th];  {\it Scalar fields on $\lambda$-deformed cosets}, Nucl. Phys.{\bf B938} (2019) 671, arXiv: 1808.02971 [hep-th]}

  \bibitem{LO}{D. L\"ust and D. Osten, {\it Generalised fluxes, Yang-Baxter deformations and the $O(d,d)$ structure of non-abelian T-duality}, JHEP {\bf 1805} (2018) 165,  arXiv:1803.03971 [hep-th] }
  
  \bibitem{MY}{T. Matsumoto and K. Yoshida, {\it Lunin-Maldacena backgrounds froom the classical Yang-Baxter equation - Towards the gravity/CYBE correspondence}, JHEP {\bf 1406} (2014) 135,
  arXiv:1404.1838 [hep-th]}
  
  \bibitem{Moh}{N. Mohammedi, {\it On the geometry of classical integrable  two-dimensional nonlinear $\sigma$-models}, {\it Nucl.Phys.} {\bf B839} (2010) 420, arXiv:0806.0550 [hep-th]}
  
  \bibitem{NR}{R. Negr\'on and V. Rivelles, {\it Yang-Baxter deformations of the $AdS\times \mathbb CP3$ superstring sigma model}, 
 JHEP {\bf 1811} (2018) 043,
  arXiv:1809.01174 [hep-th]}
  
 \bibitem{ORSY} {D. Orlando, S. Reffert, J. Sakamoto and K. Yoshida, 
{\it Generalized type IIB supergravity equations and non-Abelian classical r-matrices},  J.Phys. {\bf A49} (2016) no.44, 445403, 
 arXiv:1607.00795 [hep-th]}
 
 \bibitem{OV}{D. Osten and S.J. van Tongeren, {\it Abelian Yang-Baxter deformations and TsT transformations}, Nucl. Phys. {\bf B915} (2017) 184-205, arXiv:1608.08504  [hep-th]}

   \bibitem{OQ92} {X. de la Ossa and F. Quevedo, {\it Duality symmetries from nonAbelian isometries in string theory}, Nucl.Phys. {\bf B403} (1993) 377-394, hep-th/9210021}
  
  \bibitem{SS}{J. Sakamoto and Y. Sakatani, {\it  
   Local $\beta$-deformations and Yang-Baxter sigma model},
  JHEP {\bf 1806} (2018) 147,
 arXiv:1803.05903 [hep-th]} 
   
   \bibitem{SUY}{Y. Sakatani, S. Uezara and K. Yoshida, {\it Generalized gravity from modified DFT},
  JHEP {\bf 1704} (2017) 123, 
 arXiv:1611.05856 [hep-th]}
 
  \bibitem{SV}{P. \v Severa and F. Valach, {\it Courant algebroids, Poisson-Lie T-duality, and type II supergravities}, arXiv:1810.07763 [math.DG] }

\bibitem{S98}{ K. Sfetsos, {\it Duality invariant class of two-dimensional field theories} Nucl. Phys. {\bf B561},   (1999) 316, hep-th/9904188; {\it Poisson-Lie T duality beyond the classical level and the renormalization group},
  Phys.Lett. {\bf B432} (1998) 365-375, hep-th/9803019}


  


    


 \bibitem{S14} {
   K.~Sfetsos,
  {\it Integrable interpolations: From exact CFTs to non-Abelian T-duals}, 
  Nucl.\ Phys.\ B {\bf 880} (2014) 225, 
  arXiv:1312.4560 [hep-th]}
  
   
 \bibitem{SST15} {K. Sfetsos, K. Siampos and D. Thompson,
 {\it Generalised integrable $\lambda$-  and $\eta$-deformations and their relation},
  Nucl.Phys. {\bf B899} (2015) 489-512,
  arXiv:1506.05784 [hep-th]}
  
   \bibitem{SfS}{K. Sfetsos and K. Siampos, {\it Quantum equivalence in Poisson-Lie T-duality},
 JHEP {\bf 0906} (2009) 082,
 arXiv:0904.4248 [hep-th]}
  \bibitem{SSi}{} K.~Sfetsos and K.~Siampos,
  {\it The anisotropic $\lambda$-deformed SU(2) model is integrable},
  Phys.\ Lett. {\bf B743} (2015) 160, arXiv:1412.5181 [hep-th].   
 \bibitem{SSD10}{K. Sfetsos, K. Siampos and D. Thompson, {\it Renormalization of Lorentz non-invariant actions and manifest T-duality}, Nucl.Phys. {\bf B827} (2010) 545-564, arXiv:0910.1345 [hep-th]}
  
    \bibitem{S99} {A. Stern, {\it 
  T duality for coset models},
  Nucl.Phys. {\bf B557} (1999) 459-479,
  hep-th/9903170 }
  
   \bibitem{T}{A.A. Tseytlin, {\it On a 'Universal' class of WZW type conformal models}, Nucl. Phys. {\bf B418} (1994) 173, hep-th/9311062}
  
   \bibitem{VKS}{G. Valent, C. Klim\v c\'\i k and R. Squellari,
{\it  One loop renormalizability of the Poisson-Lie sigma models},
  Phys.Lett. {\bf B678} (2009) 143-148, arXiv:0902.1459 [hept-th]}
  
  \bibitem{VT}{S.J. van Tongeren, {\it On classical Yang-Baxter based deformations of the $AdS_5\times S^5$  superstring}, JHEP {\bf 1506} (2015) 048, arXiv:1504.05516 [hep-th]; {\it 
On Yang-Baxter models, twist operators, and boundary conditions},
  J.Phys. {\bf A51} (2018) no.30, 305401,
  arXiv:1804.05680 [hep-th]}
  
  
  
  \bibitem{V} {B. Vicedo, {\it On integrable field theories as dihedral affine Gaudin models}, to appear in Int. Math. Res. Not.,
  arXiv:1701.04856 [hep-th] }
  

 
 


\end{thebibliography}
\end{document}